\documentclass[aps,reprint,twocolumn,superscriptaddress,showpacs]{revtex4-1}
\usepackage{graphicx}
\usepackage{mathrsfs}
\usepackage{bm}
\usepackage{amsmath}
\usepackage{dcolumn}
\usepackage{epstopdf}
\usepackage{dsfont}
\usepackage{amssymb}
\usepackage{tabularx}
\usepackage{array,diagbox}
\usepackage{float}
\usepackage{color}
\usepackage{epstopdf}
\usepackage{mathrsfs}
\usepackage[colorlinks=true, letterpaper=true, pdfstartview=FitV,
  linkcolor=blue, citecolor=blue, urlcolor=blue]{hyperref}

\begin{document}

\title{Semiclassical Dynamics and Nonlinear Charge Current}

\author{Yang Gao}
\affiliation{Department of Physics, Carnegie Mellon University, Pittsburgh, PA 15213, USA}

\begin{abstract}
Electron conductivity is an important material property that can provide a wealth of information about the underlying system. Especially, the response of the conductivity with respect to electromagnetic fields corresponds to various nonlinear charge currents, which have distinct symmetry requirements and hence can be used as efficient probes of different systems. To help the band-structure engineering of such nonlinear currents, a universal treatment of electron dynamics up to second order expressed in the basis of the unperturbed states are highly useful. In this work, we review the general semiclassical framework of the nonlinear charge currents.
\end{abstract}

\maketitle

\section{Introduction}

Electron transport has long been a central topic in solid state physics, not only due to its fascinating complexity but also because it can provide rich information about the underlying electronic structure and dynamics. In the most simple Drude theory of metals, the conductivity is determined by the density, effective mass of carriers, and the phenomenological relaxation time~\cite{Mermin1976}. Such simple picture can be generalized using the dynamics of Bloch electrons, to take into account the real energy spectrum and the Fermi surface geometry, as well as the microscopic scattering process~\cite{Mermin1976}.

In recent years, this classic paradigm of electron transport receives profound modifications, due to a deeper understanding of the geometric structure in the momentum space induced by the wave functions instead of the energy spectrum. The most well-known examples are various Hall effects~\cite{Nagaosa2010,Xiao2010}. Central to the Hall effect is the concept of the Berry curvature. Its integration yields the Chern number, responsible for a topological classification of matter~\cite{Qi2011}. Moreover, it directly modifies the electron motion and hence is indispensable in describing the static and optical transport phenomenon~\cite{Xiao2010,Morimoto2016,Zhong2016,Ma2015}. Berry curvature is also the analog of the magnetic field in the momentum space. As such, the source of Berry curvature is the momentum-space analog of magnetic monopoles, which can be realized by Weyl/Dirac points~\cite{Son2013,Armitage2018}. In three dimensions, such isolated monopoles in momentum space can response differently to electromagnetic fields and cause a characteristic strong negative longitudinal magnetoresistance~\cite{Son2013,Armitage2018}.

On the other hand, the charge current can be expanded with respect to the electromagnetic fields as follows
\begin{equation}\label{eq_curexp}
J_i=\sigma_{ij}E_j+\sigma_{ijk}E_jE_k+\sigma_{(ij,k)}E_jB_k+\sigma_{(ij,kk)}E_jB_kB_k+\cdots\,.
\end{equation}
Berry curvature offers a first order correction to the electron dynamics. As such, it can give an accurate description of $\sigma_{ij}$ whose antisymmetric part is the anomalous Hall effect. In addition to this linear current, there are various nonlinear currents with coefficients satisfying different symmetry requirements and hence can be used to probe distinct systems. To further demonstrate the symmetry requirement, we can expand the last three coefficients in terms of transport relaxation time $\tau$~\footnote{the highest order of $\tau$ is the same as the highest order of the field, which is the property of the asymptotic solution to the Boltzmann equation, as implied in Eq.~\eqref{eq_dis}.}
\begin{align}
\label{eq_jee}\sigma_{ijk}&=\sigma_{ijk,0}+\sigma_{ijk,1}\tau+\sigma_{ijk,2}\tau^2\,,\\
\label{eq_jeb}\sigma_{(ij,k)}&=\sigma_{(ij,k),0}+\sigma_{(ij,k),1}\tau+\sigma_{(ij,k),2}\tau^2\,,\\
\label{eq_jebb}\sigma_{(ij,kk)}&=\sigma_{(ij,kk),0}+\sigma_{(ij,kk),1}\tau+\sigma_{(ij,kk),2}\tau^2+\sigma_{(ij,kk),3}\tau^3\,.
\end{align}
The symmetry of different coefficients is summarized in Table.~\ref{tb_tb1}. It is thus tempting to find how different coefficients depend on the geometric properties of Bloch electrons in the momentum space. This can provide experimental guide and help the band structure engineering of larger nonlinear charge currents. For this purpose, the electron dynamics should be extended to second order.

\begin{table}[t]
\label{tb_tb1}
\centering
\caption{Nonzero components of conductivity in crystals with different space inversion and time reversal symmetries. The requirement of the lower right entry should be understood as in crystals that simultaneously break time reversal and inversion symmetry but preserve the combined symmetry.}
\begin{ruledtabular}
\begin{tabular}{l|c|l}
\diagbox{Time}{Space}& Even & Odd\\
\hline
Even & $\begin{tabular}{c}
$\sigma_{(ij,k),0}$\quad$\sigma_{(ij,k),2}$\\
 $\sigma_{(ij,kk),1}$ \quad  $\sigma_{(ij,kk),3}$
\end{tabular}$
& $\sigma_{ijk,1}$\\
\hline
Odd & $\begin{tabular}{c}
\quad$\sigma_{(ij,k),1}$ \\
$ \sigma_{(ij,kk),0}$\quad  $\sigma_{(ij,kk),2}$
\end{tabular}$
& $\sigma_{ijk,0}$\quad $\sigma_{ijk,2}$
\end{tabular}
\end{ruledtabular}
\end{table}

In this work, we will focus on the semiclassical treatment of the electron dynamics and review the effort of generalizing such theory up to second order and deriving the nonlinear currents in the Boltzmann transport framework. Our paper is organized as follows. In Sect.~II, we discuss the generalization of the semiclassical equations of motion up to second order through an appropriately constructed wave packet up to first order. We then discuss the validity of the semiclassical theory though the magnetoelectric coefficient and magnetic susceptibility. With the correct semiclassical theory up to second order, in Sect.~III, we extend the Boltzmann transport theory and derive its solution in steady-state up to third order. We then use this universal treatment to derive various nonlinear currents, including the electric-field-correction to the nonlinear Hall effect~($\sigma_{jii}$), linear magnetoresistance~($\sigma_{(ij,k)}$), and quadratic magnetoresistance~($\sigma_{(ij,kk)}$).

\section{Semiclassical Theory of Bloch electrons in electromagnetic fields}
To derive various nonlinear charge currents, one will need the correct electron dynamics beyond linear order under electromagnetic fields. For this purpose, certain perturbation technique should be employed. However, the periodic nature of crystals causes difficulties for such process. We note that the perturbation from electromagnetic fields reads $e\bm E\cdot \bm r+\frac{1}{2}e[\bm A(\bm r)\cdot \hat{\bm v}+\hat{\bm v}\cdot \bm A(\bm r)]$. It contains the position operator, which is unbounded in extended systems. Therefore, direct application of the time-independent perturbation is inappropriate.

To circumvent such difficulty, for magnetoconductivty one can first obtain the eigenstate of Bloch electrons under magnetic field, which are Landau levels for effective continuum models or Hofstadter spectrum for tight-binding models. Then the static magnetoconductivity is obtained using the standard Kubo formula under an appropriate limit. The dependence of the conductivity on magnetic field is encoded in the quantization of original Bloch states under magnetic field. Although this method is standard and has wide applicability, it cannot give universal understanding of the physics behind the magnetoconductivity, as the quantization into Landau levels or the Hofstadter spectrum varies for different systems.

Alternatively, one can construct a phase space using an appropriate position variable $\bm R$ and momentum variable $\bm P$ from the Bloch states. The electron dynamics is then contained in the equations of motion in the phase space. If $\bm R$ and $\bm P$ are canonical variables, their dynamics is solely determined by the energy. Therefore, the whole procedure amounts to deriving an effective energy in terms of $\bm R$ and $\bm P$ under external fields. If $\bm R$ and $\bm P$ are non-canonical, their Poisson brackets are also needed, which determine the structure of the equations of motion. The magnetoconductivity is then derived in combination with the Boltzmann equation for the distribution function. Although this method treats external fields as perturbations, and is only derived up to the second order currently, it has the benefit that it yields analytical results independent of the model detail. In fact, all terms in the equations of motion are expressed using the unperturbed Bloch functions, and hence can directly access both the spectrum and the geometric structure in the unperturbed Hilbert space.

In this section, we follow the second method by treating Bloch electrons as wave packets, whose center of mass position and momentum naturally span the phase space. We will first sketch the construction of the wave packet and show how its internal structure modifies the electron dynamics in phase space at first order of external fields. A detailed discussion of both aspects can be found in Ref.~\cite{Xiao2010}. We then extend such theory up to second order and derive the second order correction to equations of motion. The validity of the semiclassical theory can be confirmed through the magnetoelectric effect and the quantization of effective energies into Landau levels under magnetic field. Since all elements in the semiclassical dynamics are expressed using unperturbed Bloch states, they can be evaluated using the output from the first-principles codes.

\subsection{Properties of the wave packet}
The dynamics of Bloch electrons under uniform electromagnetic fields is governed by the following crystal Hamiltonian in non-relativistic limit
\begin{equation}
\hat{H}_\text{f}=\hat{H}_0[\hat{\bm p}+e\bm A(\bm r);\bm r]+\frac{g\mu_B}{\hbar}\bm B\cdot \hat{\bm s}+e\bm E\cdot \bm r\,,
\end{equation}
where $\hat{H}_0(\hat{\bm p};\bm r)$ is the unperturbed Hamiltonian for periodic crystals with $\hat{\bm p}$ being the momentum operator, $g$ is the gyromagnetic ratio, $\mu_B$ is the Bohr magneton, and $\hat{\bm s}$ is the spin operator. The magnetic field enters through both the minimal coupling and the Zeeman coupling. Generally speaking, $\hat{H}_\text{f}$ does not respect the translational symmetry, unless the magnetic flux through a unit cell is a rational number times the flux quantum such that the magnetic translational symmetry is recovered~\cite{Zak1964a, Zak1964b,Hofstadter1976,Chang1994}.

Although $\hat{H}_\text{f}$ exactly determines the dynamics, to directly diagonalize it is a difficult task. Instead, we can simplify the problem by assuming that the solution to the Schr$\ddot{\rm o}$dinger equation $(i\hbar \partial_t-\hat{H}_\text{f})\psi=0$ takes the ansartz of a wave packet, which is the essence of the semiclassical theory. To construct the wave packet, we first evaluate the electromagnetic potential in the full Hamiltonian at the center of mass position $\bm r_c$ of the wave packet, and hence recover the translational symmetry of the original unperturbed crystal. The resulting local Hamiltonian reads:
\begin{equation}
\hat{H}_c=\hat{H}_0[\hat{\bm p}+e\bm A(\bm r_c);\bm r]+\frac{g\mu_B}{\hbar}\bm B\cdot \hat{\bm s}+e\bm E\cdot \bm r_c\,,
\end{equation}
 Its eigenenergy forms the local Bloch bands $\varepsilon_n[\bm p+(e/\hbar)\bm A(\bm r_c)]+e\bm E\cdot \bm r_c$ with $n$ being the band index and $\hbar\bm p$ being crystal momentum. Its eigenfunction is the Bloch function $e^{i\bm p\cdot \bm r}|u_n[\bm p+e\bm A(\bm r_c)/\hbar]\rangle$.

In the following, we focus on a single band with index $0$ and assume that it is well separated from all the other bands. When electromagnetic fields are weak, the dynamics of a Bloch electron starting from some state in this band will still be confined in the same band. Therefore, the wave packet is constructed as the superposition of Bloch states from the band $0$~\cite{Sundaram1999}
\begin{equation}
|W\rangle=\int d\bm p C_0(\bm p) e^{i\bm p\cdot \bm r}|u_0[\bm p+e\bm A(\bm r_c)/\hbar]\rangle\,.
\end{equation}
$|W\rangle$ should be normalized, indicating $\int d\bm p |C_0|^2=1$.

There are two constraints on the coefficient $C_0$. First, the wave packet is assumed to be sharply localized in the momentum space. This requires that the magnitude of $C_0$ satisfies $|C_0|^2\approx \delta (\bm p-\bm p_c)$. Here $\hbar\bm p_c$ is the center of mass momentum of the wave packet, i.e.
\begin{equation}
\bm p_c=\langle W|\bm p|W\rangle\,.
\end{equation}
Secondly, the construction of the wave packet requires a local Hamiltonian around the center of mass position $\bm r_c$. This $\bm r_c$ has to be determined in a self-consistent manner, which exerts a constraint on the phase of $C_0$~\cite{Sundaram1999}:
\begin{equation}\label{eq_wrc}
\bm r_c=\langle W|\bm r|W\rangle=\left.\frac{\partial \gamma}{\partial \bm p}\right |_{\bm p=\bm p_c}+ \bm {\mathcal{A}}_0(\bm k_c)\,,
\end{equation}
where $\gamma=-arg(C_0)$, $\bm {\mathcal{A}}_0(\bm p)=\langle u_0(\bm p)|i\bm \partial_{\bm p}|u_0(\bm p)\rangle$ is the intraband Berry connection, and $\hbar\bm k_c=\hbar\bm p_c+e\bm A(\bm r_c)$ is the gauge-independent physical momentum. These two constraints complete the construction of the wave packet.
Although the wave packet is for Bloch states from a single band, it can be generalized to account for the general multiband case~\cite{Culcer2005,Shindou2005}.

\subsection{Semiclassical dynamics up to first order}
With the knowledge of the wave packet, we now derive its evolution. The coefficient $C_0$ in the wave packet can be determined through the variational principle, i.e. the least action principle. Since $C_0$ is a complex function, with independent magnitude and phase, determined by $\bm k_c$ (or equivalently, $\bm p_c$) and $\bm r_c$ respectively, applying the variational principle to the Lagrangian will yield the dynamics of $\bm r_c$ and $\bm k_c$, i.e. the phase space equations of motion.

By evaluating the Lagrangian under the wave packet up to first order, we obtain~\cite{Sundaram1999}
\begin{align}\label{eq_lag}
L&=\langle W|i\hbar\partial_t-\hat{H}_0-\hat{H}_1|W\rangle\notag\\
&=-(\bm r_c-\bm {\mathcal{A}}_0)\cdot \hbar\dot{\bm k}_c-\frac{1}{2}e\bm B\times \bm r_c\cdot \dot{\bm r}_c-\tilde{\varepsilon}_0\,,
\end{align}
where $\hat{H}_1=\frac{e}{4} \bm B\cdot [(\bm r-\bm r_c)\times \hat{\bm v}-\hat{\bm v}\times (\bm r-\bm r_c)]+e\bm E\cdot (\bm r-\bm r_c)$ is the first order correction to the local Hamiltonian $\hat{H}_c$ and $\tilde{\varepsilon}=\varepsilon_0-\bm B\cdot \bm m+e\bm E\cdot \bm r_c$ with $\bm m=\bm m_{orb}-(g\mu_B/\hbar)\langle u_0|\hat{\bm s}|u_0\rangle$ is the modified Band energy after taking account of both
the magnetic moment and electric dipole of the wave packet, which will be discussed later. Here we use the symmetric gauge for the vector potential.

The Berry connection $\bm {\mathcal{A}}_0$ is an essential ingredient in the Lagrangian. In the absence of $\bm {\mathcal{A}}_0$, the above Lagrangian reduces to the conventional canonical form for electrons under electromagnetic fields:
\begin{equation}
L=\left(\hbar \bm k_c-\frac{1}{2}e\bm B\times \bm r_c\right)\cdot \dot{\bm r}_c-\tilde{\varepsilon}_0\,.
\end{equation}
The appearance of $\bm {\mathcal{A}}_0$ indicates that $\bm r_c$ and $\bm k_c$ are not canonical variables. To make this statement clearer, we make the following transformation~\cite{Chang2008}
\begin{align}
\label{eq_rc}\bm r_c&=\bm q+\bm {\mathcal{A}}_0+\frac{1}{2}\frac{e}{\hbar}(\bm B\times \bm {\mathcal{A}}_0\cdot \bm \partial_{\bm p})\bm {\mathcal{A}}_0+\frac{1}{2}\frac{e}{\hbar}\bm \Omega_0\times (\bm B\times \bm {\mathcal{A}}_0)\,,\\
\label{eq_kc}\bm k_c&=\bm p+\frac{1}{2}\frac{e}{\hbar}\bm B\times \bm q+\frac{e}{\hbar}\bm B\times (\bm r_c-\bm q)\,,
\end{align}
where $\bm \Omega_0=\bm \nabla_{\bm p}\times \bm {\mathcal{A}}_0$ is the Berry curvature, which can be viewed as the geometrical Berry phase per unit area~\cite{Berry1984}.
Here the argument of $\bm {\mathcal{A}}_0$ and $\bm \Omega_0$ is $\bm p+\frac{e}{2\hbar}\bm B\times \bm q$. Then the Lagrangian in Eq.~\eqref{eq_lag} recovers the canonical form for $\bm p$ and $\bm q$
\begin{equation}
L=\hbar\bm p\cdot \dot{\bm q}-\tilde{\varepsilon}_0\,.
\end{equation}
Eq.~\eqref{eq_rc} and \eqref{eq_kc} thus describe the connection between physical variables and canonical variables.

One consequence of the noncanonicality is that the phase space measure for the volume element $d\bm r_cd\bm k_c$ has to change based on the Jacobian $\partial(\bm r_c,\bm k_c)/\partial(\bm q,\bm p)$. The resulting phase space density of states reads~\cite{Xiao2005}
 \begin{equation}
 \mathcal{D}=1+\frac{e}{\hbar}\bm B\cdot \bm \Omega_0(\bm k_c)\,.
  \end{equation}
Since the physical position and momentum take the crystal volume and the Brillouin zone as their range, it is natural to evaluate the statistical average of any operator in the physical phase space spanned by $(\bm r_c,\bm k_c)$. Therefore, the modified density of states $\mathcal{D}$ should always be used if the magnetic field is involved.

The noncanonicality of $\bm r_c$ and $\bm k_c$ will also lead to a nontrivial symplectic form, and hence affects the structure of the equations of motion. From Eq.~\eqref{eq_lag}, the Euler-Lagrangian equations of motion yield the following phase space dynamics~\cite{Sundaram1999}
\begin{align}
\label{eq_dotrc}\dot{\bm r}_c&=\frac{\partial \tilde{\varepsilon}_0}{\hbar\partial \bm k_c}-\dot{\bm k}_c\times \bm \Omega_0(\bm k_c)\,,\\
\label{eq_dotkc}\hbar\dot{\bm k}_c&=-e\bm E-e\dot{\bm r}_c\times \bm B\,.
\end{align}
The second term in the velocity equation is the anomalous velocity arising from the noncanonicality. The above equations of motion have wide applicability in the static spin, heat, and charge transport phenomena~\cite{Xiao2010} and optical phenomena~\cite{Moore2010,Ma2015,Zhong2015,Morimoto2016,Zhong2016,Morimoto2016a,Deyo2019}.

\subsection{Electromagnetic Dipoles}
One direct application of the semiclassical theory is to derive the electromagnetic multipoles in crystals. This derivation confirms the validity of the semiclassical theory as the results are consistent with those derived using other methods. As the wave packet is sharply localized in the momentum space, it has a finite width in the real space. The charge and current distributions across such spread need not to be uniform. To represent such internal anisotropy of the wave packet, it is natural to use electromagnetic multipoles. For illustration purpose, we will consider the electromagnetic dipoles.

The electric dipole of the wave packet reads $\langle W|-e\bm r|W\rangle=-e\bm r_c$, with $\bm r_c$ given in Eq.~\eqref{eq_wrc}. Under the periodic gauge~\cite{Resta2000,Xiao2010}, the Bloch wave function is periodic in the momentum space, so is the coefficient $C_0$. Therefore, by integrating $\bm r_c$ in the Brillouin zone, the first term in Eq.~\eqref{eq_wrc} from the phase of $C_0$ can at most yield an integer multiples of $2\pi$ and hence can be ignored. The remaining term yields the electric polarization in crystals
\begin{equation}\label{eq_dipole}
\bm P=-e\int \frac{d\bm k_c}{(2\pi)^3} \bm {\mathcal{A}}_0(\bm k_c)\,.
\end{equation}
Originally, the electric polarization is derived by first considering the adiabatic charge pumping current and then identifying it as the displacement current from the temporal variation of the polarization~\cite{Kingsmith1993,Resta1994}. Here our method yields the same electric polarization under the periodic gauge for Bloch functions.

The magnetic dipole can be evaluated as follows~\cite{Chang1996,Sundaram1999}
\begin{align}\label{eq_m}
\bm m_\text{orb}&=-e\langle W|\frac{1}{2}[(\bm r-\bm r_c)\times\hat{\bm v}-\hat{\bm v}\times (\bm r-\bm r_c)]|W\rangle\notag\\
&=-\frac{e}{2} \sum_{n\neq 0}\bm {\mathcal{A}}_{0n}\times \bm v_{n0}\,,
\end{align}
where $\bm {\mathcal{A}}_{0n}=\langle u_0|i\bm \partial_{\bm k_c}|u_n\rangle$ is the interband Berry connection and $\bm v_{n0}=\langle u_n|\hat{\bm v}|u_0\rangle$ is the interband velocity element. In this work, we use the convention that for intraband matrix elements, we only keep one band index. The magnetic moment $\bm m_\text{orb}$ represents how a wave packet couples to an external magnetic field, i.e. shifting the band energy through an effective Zeeman coupling $-\bm B\cdot \bm m_\text{orb}$. As such, $\bm m_\text{orb}$ contributes to the orbital magnetization as shown later. $\bm m_\text{orb}$ can usually be accessed through the optical activity of crystals. For example, in ferromagnetic materials, the cross-gap part of $\bm m_\text{orb}$ (with $0$ standing for occupied bands and $n$ for the unoccupied bands in Eq.~\eqref{eq_m}) can be measured through the $f$-sum rule for the circular dichroism~\cite{Souza2008,Yao2008}. Moreover, in noncentrosymmetric metals, the natural optical activity in the semiclassical regime is determined by the integration of the momentum space dipole of the magnetic moment (defined as $v_i (m_\text{orb})_j$ with $\bm v$ being the band velocity) on the Fermi surface~\cite{Ma2015,Zhong2015,Zhong2016}.

Interestingly, if one starts from the relativistic Dirac Hamiltonian and construct a coherent wave packet for the upper two bands, the orbital magnetic moment of this wave packet naturally recovers the spin magnetic moment~\cite{Chang2008}. Mathematically, this is equivalent to the Foldy-Wouthuysen transformation that reduces the Dirac Hamiltonian to the non-relativistic Schr$\ddot{\rm o}$dinger Hamiltonian~\cite{Foldy1950,Blount1962}. Here the semiclassical theory offers a heuristic and clear picture for the emergence of spin.

 To derive the orbital magnetization in crystals, one should evaluate the free energy under magnetic field. On one hand, the magnetic moment of the wave packet couples to the magnetic field and modifies the band energy. On the other hand, the phase space density of states is changed by the magnetic field. Taking both corrections into consideration, one obtains the following free energy
\begin{equation}
F=\int \mathcal{D}\frac{d\bm k}{(2\pi)^3}  (-k_BT)\ln \left[1+\exp\left(\frac{\tilde{\varepsilon}_0-\mu}{k_BT}\right)\right]\,.
\end{equation}
Therefore, the orbital magnetization can be derived by taking the derivative of the free energy with respect to the magnetic field~\cite{Xiao2006}
\begin{align}
\bm M=-\frac{\partial F}{\partial B}=\int \frac{d\bm k}{(2\pi)^3} \left(f_0 \bm m-\frac{e}{\hbar}g_e\bm \Omega_0\right)\,,
\end{align}
where $g_e=-(k_BT)\ln[1+\exp((\varepsilon_0-\mu)/k_BT)]$ is the grand potential density, and $f_0$ is the equilibrium Fermi distribution function. Here and hereafter we will ignore the subscript $c$ in the integration over $\bm k_c$. This is the contribution of the orbital magnetization from band $0$. The total contribution can be obtained by summing over the band index. The orbital magnetization thus contains two contributions. The first one is due to the relative motion inside the wave packet, or the self-rotation of the wave packet. The second one is due to the rotation of the wave packet as a whole, or the revolution of the wave packet. This same magnetization can also be obtained through the Wannier function approach~\cite{Thonhauser2005,Ceresoli2006}, the exact Hofstadter spectrum~\cite{Gat2003a,Gat2003b}, or the linear response theory~\cite{Shi2007,Qin2011}.

\subsection{Semiclassical dynamics up to second order}
Although the semiclassical dynamics in Eq.~\eqref{eq_dotrc} and \eqref{eq_dotkc} is very powerful, it cannot fully account for the second order response functions such as the magnetoresistance, magnetoelectric coefficient, and so on. For this purpose, we need to extend the above semiclassical dynamics up to second order.

The construction of the wave packet has to be modified accordingly. The external electromagnetic fields generally affect the eigenfunction. As a result, the wave packet should be the superposition of the true eigenstates instead of the unperturbed ones. We can still expand the true eigenstate in the basis of the unperturbed ones, and obtain the following wave packet~\cite{Gao2014}
\begin{equation}\label{eq_wp2}
|W\rangle=\int d\bm p e^{i\bm p\cdot \bm r}\left(C_0(\bm p) |u_0\rangle+\sum_{n\neq 0}C_n(\bm p)|u_n\rangle\right)\,.
\end{equation}
Here for simplicity we drop the argument $\bm p+(e/\hbar)\bm A(\bm r_c)$ of $|u_0\rangle$ and $|u_n\rangle$.
If $|W\rangle$ satisfies the Schr$\ddot{\rm o}$dinger equation, we have
\begin{align}
\langle u_n|e^{-i\bm p\cdot \bm r}(i\hbar \partial_t -\hat{H}_\text{f})|W\rangle=0\,,\quad \forall\; n\neq 0\,.
\end{align}
Therefore, we use this as an additional constraint on the wave packet in Eq.~\eqref{eq_wp2}, which determines the connection between $C_n$ and $C_0$. Specifically, solving this constraint at the linear order of electromagnetic fields yields~\cite{Gao2014}
\begin{equation}\label{eq_cn}
C_n=\frac{G_{n0}}{\varepsilon_0-\varepsilon_n}C_0-\frac{i}{2}\frac{e}{\hbar}[\bm B\times (i\bm \partial_{\bm p}+\bm {\mathcal{A}}_0 -\bm r_c)C_0]\cdot \bm {\mathcal{A}}_{n0}\,,
\end{equation}
where $G_{n0}=-\bm B\cdot \bm M_{n0}+e\bm E\cdot \bm {\mathcal{A}}_{n0}$ with $\bm M_{n0}=\frac{e}{2}\sum_{m\neq 0} (\bm v_{nm}+\bm v_0\delta_{mn})\times \bm {\mathcal{A}}_{m0}-g\mu_B \bm s_{n0}/\hbar$ being the interband element of the total~(spin plus orbital) magnetic moment. We comment that different from the discussion here, in Ref.~\cite{Gao2014} and \cite{Gao2015}, the spin magnetic moment has not been explicitly added.

\begin{figure}
  \centering
  \includegraphics[width=\columnwidth]{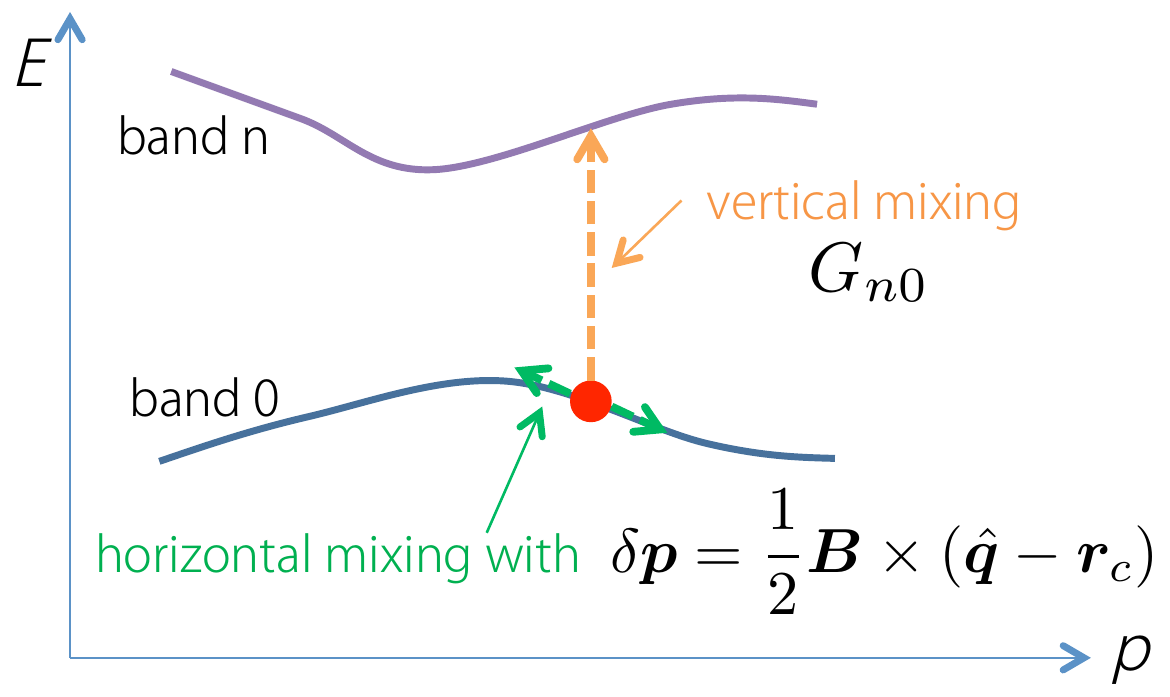}\\
  \caption{The correction to wave packet from magnetic field. $\hat{\bm q}$ has the same meaning of $\bm r$. From Ref.~\cite{Gao2015}.}\label{fig_perturb}
\end{figure}

Equation~\eqref{eq_cn} indicates that the correction to the wave packet from the magnetic field has two contributions with distinct structures and hence different origins. It is useful to first analyze the nature of these two perturbations as they prelude the form of the second order correction to the phase space dynamics. The first term has the conventional form of the perturbation correction with the energy gap as the denominator. It is called the vertical mixing as it mixes Bloch states from different bands but at the same $\bm p$ point in the Brillouin zone. It also contains both adiabatic and nonadiabatic perturbation, as the wave packet $|W\rangle$ is time-dependent through the argument $\bm p+(e/\hbar)\bm A(\bm r_c)$ of the Bloch state.

In comparison, the second term in $C_n$ modifies the wave packet in the following way:
\begin{align}\label{eq_hoz}
&\int d\bm p e^{i\bm p\cdot \bm r} \frac{-i}{2} \frac{e}{\hbar}[\bm B\times (i\bm \partial_{\bm p}+\bm {\mathcal{A}}_0-\bm r_c)C_0]\cdot \sum_{n\neq 0}\bm {\mathcal{A}}_{n0}|u_n\rangle\notag\\
=&\int d\bm p e^{i\bm p\cdot \bm r} \frac{e}{2\hbar}\left(\bm B\times (\bm r-\bm r_c)\cdot \hat{\bm D}|u_0\rangle+\bm B\cdot \bm\Omega_0 |u_0\rangle\right)\,.
\end{align}
where $\hat{D}=\bm \partial_{\bm p}+i\bm {\mathcal{A}}_0$ is the covariant derivative that ensures the gauge invariance. Notice that the momentum argument of $|u_0\rangle$ is $\bm p+(e/\hbar)\bm A(\bm r_c)$. The first term in Eq.~\eqref{eq_hoz} has the meaning of shifting the momentum to $\bm p+(e/\hbar)\bm A(\bm r)$. This is a unique property of the perturbation from the magnetic field, and it suggests that the correction to the wave function also obeys the Peierls substitution. The same property is also derived using the Moyal product and phase space formulation of quantum mechanics~\cite{Blount1962b}. The second term in Eq.~\eqref{eq_hoz} makes the total correction normal to the original state $e^{i\bm p\cdot \bm r} |u_0\rangle$ to eliminate redundancy in the perturbative correction. Since the second term in $C_n$ effectively mixes Bloch states in the same band but at neighbouring $\bm p$ points, it is referred to as the horizontal mixing. Both corrections can be easily visualized in Fig.~\ref{fig_perturb}.

Using the modified wave packet, we obtain the following Lagrangian
\begin{equation}\label{eq_lag2}
L=-(\bm r_c-\bm {\mathcal{A}}_0^\text{t})\cdot \hbar\dot{\bm k}_c-\frac{1}{2}e\bm B\times \bm r_c\cdot \dot{\bm r}_c-\tilde{\varepsilon}_0\,,
\end{equation}
where $\bm {\mathcal{A}}_0^\text{t}$ is the Berry connection evaluated in the true eigen state instead of the unperturbed one. It can be put in the form of the original Berry connection plus a positional shift correction: $\bm {\mathcal{A}}_0^\text{t}=\bm {\mathcal{A}}_0+\bm {\mathcal{A}}_0^\prime$. Before discussing the expression of the positional shift $\bm {\mathcal{A}}_0^\prime$ and modified energy $\tilde{\varepsilon}$, we comment that the Lagrangian up to second order has the exact same form with the previous one in Eq.~\eqref{eq_lag}. Consequently, the connection between physical and canonical variables also has the same form except that the Berry connection $\bm {\mathcal{A}}_0^\text{t}$ in the exact eigenstate should be used
\begin{align}
\label{eq_rc2}\bm r_c&=\bm q+\bm {\mathcal{A}}_0^\text{t}+\frac{e}{2\hbar}(\bm B\times \bm {\mathcal{A}}_0^\text{t}\cdot \bm \partial_{\bm p})\bm {\mathcal{A}}_0^\text{t}+\frac{e}{2\hbar}\bm  \Omega_0^\text{t}\times (\bm B\times \bm {\mathcal{A}}_0^\text{t})\,,\\
\label{eq_kc2}\bm k_c&=\bm p+\frac{e}{2\hbar}\bm B\times \bm q+\frac{e}{\hbar}\bm B\times (\bm r_c-\bm q)\,,
\end{align}
where $\bm {\Omega}_0^\text{t}=\bm \nabla_{\bm p}\times \bm{\mathcal{A}}_0^\text{t}$ is the Berry curvature evaluated using the exact eigenstate. As a result, the phase space density of states reads
\begin{equation}
\mathcal{D}=1+\frac{e}{\hbar}\bm B\cdot \bm {\Omega}_0^\text{t}\,.
 \end{equation}
More importantly, the equations of motion up to second order also keep the same form~\cite{Gao2014}
\begin{align}
\label{eq_dotrc2}\dot{\bm r}_c&=\frac{\partial \tilde{\varepsilon}_0}{\hbar\partial \bm k_c}-\dot{\bm k}_c\times \bm {\Omega}_0^\text{t}(\bm k_c)\,,\\
\label{eq_dotkc2}\hbar\dot{\bm k}_c&=-e\bm E-e\dot{\bm r}_c\times \bm B\,.
\end{align}

The positional shift is gauge-independent. It represents the additional shift of the wave packet center after applying external electromagnetic fields
\begin{equation}
\bm r_c=\langle W|\bm r|W\rangle=\left.\frac{\partial \gamma}{\partial \bm p}\right |_{\bm p=\bm p_c}+ \bm {\mathcal{A}}_0(\bm k_c)+\bm {\mathcal{A}}_0^\prime(\bm k_c)\,.
\end{equation}
The positional shift at first order reads
\begin{equation}
\bm {\mathcal{A}}_0^\prime=2\sum_{n\neq 0}{\rm Re}\frac{\bm {\mathcal{A}}_{0n} G_{n0}}{\varepsilon_0-\varepsilon_n}-\frac{1}{2}\frac{eB}{\hbar}\epsilon_{ijk} \gamma_{ji\ell}\hat{e}_\ell\,,
\end{equation}
where $\epsilon_{ijk}$ is the Levi-Civita symbol and $\gamma_{ji\ell}=\frac{1}{2}(\partial_\ell g_{ij}+\partial_{i}g_{j\ell}-\partial_jg_{i\ell})$ is the Christoffel symbol. Here and hereafter, we use the Einstein summation convention for repeated indices and the partial derivative is with respect to the crystal momentum unless otherwise specified. $g_{ij}$ is a Fubini-Study metric tensor, usually called quantum metric~\cite{Provost1980,Neupert2013,Anandan1990,Resta2011}. It measures the distance between two neighbouring Bloch states in the momentum space. In fact, it is a part of a more general concept, the quantum geometrical tensor, which, for band $0$, is defined as
\begin{equation}
\mathcal{G}_{ij}=\langle \partial_{i}u_0|\partial_{j}u_0\rangle-(\mathcal{A}_{i})_0(\mathcal{A}_{j})_0\,.
\end{equation}
The real and imaginary part of $\mathcal{G}_{ij}$ yield the quantum metric and Berry curvature, respectively~\cite{Provost1980,Neupert2013}.

In the positional shift, the first term is due to the vertical mixing and the second term is due to the horizontal mixing. It is interesting that the horizontal mixing yields a purely geometric correction to the Berry connection. This is not coincidental, as the horizontal mixing addresses neighbouring Bloch states in the momentum space, whose difference is proportional to the Berry connection and whose distance is determined by the quantum metric. We comment that this positional shift is also envisioned in Ref.~\cite{Berry1987}, using a similar technique.

The positional shift reflects the change in the electric dipole moment of the wave packet. When external fields are applied, the electric polarization has to pick up the following change. First, the density of states has to be changed to $\mathcal{D}$. Second, the Berry connection $\bm {\mathcal{A}}_0\rightarrow \bm {\mathcal{A}}_0^\text{t}+\frac{e}{2\hbar}(\bm B\times \bm {\mathcal{A}}_0^\text{t}\cdot \bm \partial_{\bm p})\bm {\mathcal{A}}_0^\text{t}+\frac{e}{2\hbar}\bm { \Omega}_0^\text{t}\times (\bm B\times \bm{\mathcal{A}}_0^\text{t})$ according to Eq.~\eqref{eq_rc2}. As a result, the first order change in electric polarization reads
\begin{equation}
\bm {\delta P}=-e\int \frac{d\bm k}{(2\pi)^3} \left[\frac{e}{2\hbar}(\bm \Omega_0\cdot \bm {\mathcal{A}}_0)\bm B+\bm {\mathcal{A}}_0^\prime\right]\,.
\end{equation}
The first term is the Abelian Chern-Simons 3-form, which yields the topological part of the orbital magnetoelectric coefficient. The second term yields the electric polarizability and the cross-gap part of the orbital magnetoelectric polarizability, consistent with the calculation using the linear response theory~\cite{Essin2010}. Such consistency confirms the validity of the first order correction to the Berry phase.

The modified band energy in Eq.~\eqref{eq_lag2} should contain the correction up to second order, i.e. it can be put in the following form: $\tilde{\varepsilon}_0=\varepsilon_0-\bm B\cdot \bm m+e\bm E\cdot \bm r_c+\varepsilon_{(2)}$. The second order correction describes the change in the electric dipole and magnetic moment of the wave packet in response to external fields. It reads~\cite{Gao2015}
\begin{align}\label{eq_eng2}
\varepsilon_{(2)}=&\frac{1}{ 4}\frac{e}{\hbar} (\bm B\cdot \bm{\Omega}_0) (\bm B\cdot \bm m)-{1\over 8}\frac{e^2}{\hbar^2} \epsilon_{s i k} \epsilon_{t j \ell}B_{s} B_{t}g_{ij} \alpha_{k\ell} \notag\\
&+\bm \nabla\cdot\bm P_\text{E}+\sum_{n\neq 0}{G_{0n} G_{n0}\over \varepsilon_{0}-\varepsilon_{n}}\notag\\
&-e\bm B\cdot (\bm {\mathcal{A}}_0^\prime \times \bm v_0)+{e^2\over 8m} (B^2 g_{ii}-B_i g_{ij} B_j),
\end{align}
where $\alpha_{k\ell}=\partial_k \partial_\ell \varepsilon_0/\hbar^2$ is the inverse effective mass tensor, $m$ is the free electron mass, $G_{0n}=(G_{n0})^\star$, and
\begin{align}
\bm P_\text{E}=\frac{1}{2}e \sum_{n\neq 0}{\rm Re}[(\bm B\times \bm {\mathcal{A}}_{0n})G_{n0}]\,.
\end{align}
In the energy correction, the first two terms are purely due to the horizontal mixing and hence are geometrical corrections involving Berry curvature and quantum metric. The third term is due to the cross effect of the horizontal and vertical mixing. $\bm P_\text{E}$ is called energy polarization as it accounts for the energy dipole in the momentum space. The fourth term is purely due to the horizontal mixing and has the conventional form of the perturbation correction to the energy. The fifth term shifts the momentum argument of $\varepsilon_0$. The last term is the expectation of the second order correction to the local Hamiltonian.

We comment that if one starts from an effective tight-binding Hamiltonian, the last term of Eq.~\eqref{eq_eng2} will change. Instead of the free electron mass, one has to use the Hessian operator $\hat{\Gamma}_{ij}=\partial^2 \hat{H}_0/\partial \hat{p}_i\partial \hat{p}_j$. Then the last term in Eq.~\eqref{eq_eng2} should be replaced by the following two terms
\begin{align}\label{eq_hessian}
\varepsilon_\text{H}=&-{e^2\over 16} (\bm B\times \bm \partial)_i (\bm B\times \bm \partial)_j \langle u_0|\hat{\Gamma}_{ij}|u_0\rangle\notag\\
&+{e^2\over 8} \sum_{(m,n)\neq 0} (\bm B\times \bm {\mathcal{A}}_{0m})_i (\hat{\Gamma}_{ij})_{mn} (\bm B\times \bm {\mathcal{A}}_{n0})_j\,.
\end{align}
Here $(\hat{\Gamma}_{ij})_{mn}=\langle u_m|\hat{\Gamma}_{ij}|u_n\rangle$.

\subsection{Landau level quantization and magnetic susceptibility}
If one starts from a continuum tight-binding Hamiltonian, electronic states will be quantized into Landau levels under magnetic field. Therefore, as long as the semiclassical theory can successfully reproduce the Landau level quantization up to second order, it is enough to account for various response functions at second order in magnetic field.

There are various quantization rules for Landau levels, such as the Onsager's rule~\cite{Onsager1952}, the Maslov canonical operator method~\cite{Reijnders2013}, the Gutzwiller trace formula~\cite{Gutzwiller1971}, and so on. For semiclassical theory, the most relevant one is the Onsager's rule~\cite{Onsager1952}, which reads
\begin{align}
S(\varepsilon_n)=2\pi\left(n+{1\over 2}\right) {eB\over \hbar}\,,
\end{align}
where $S(\varepsilon_n)$ is the area in the momentum space enclosed by an equal-energy contour. Later, this quantization rule is generalized up to first order to take account of corrections from Berry phase and magnetic moment~\cite{Wilkinson1984,Rammal1985,Mikitik1999,Chang1996,Carmier2008,Xiao2010,Fuchs2010}.

\begin{figure}
  \centering
  \includegraphics[width=\columnwidth]{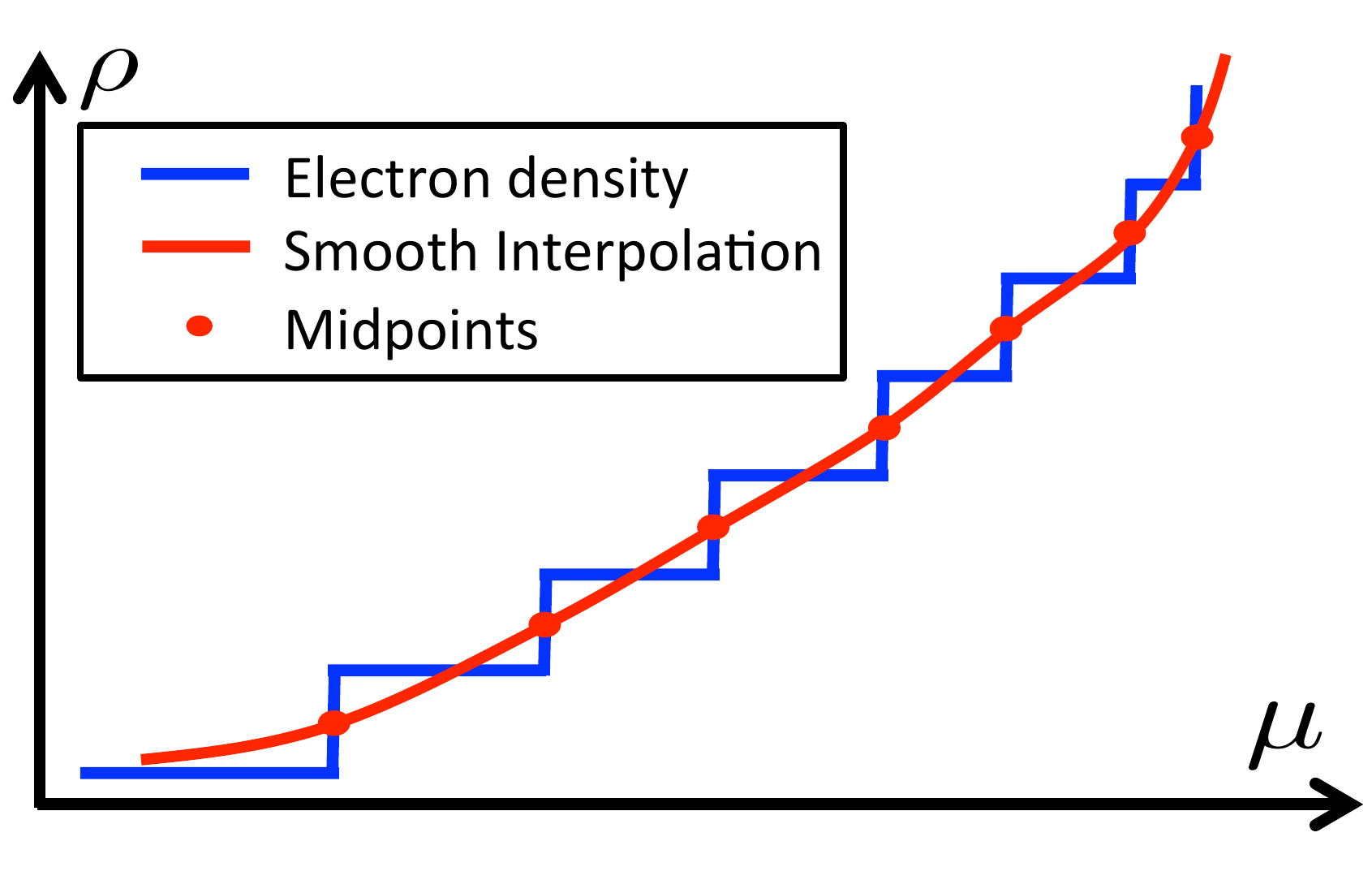}\\
  \caption{Density quantization rule for Landau levels. The true electron density in the blue line experiences a jump at each Landau level. From Ref.~\cite{Gao2017a}.}\label{fig_ll}
\end{figure}

To further generalize the Onsager's rule, one observes that the area $S(\varepsilon_n)$ is proportional to the total electron density below $\varepsilon_n$. Since the area $S$ as a function of energy is a smooth function, such electron density should be understood as the smooth semiclassical electron density $\rho_\text{semi}$. Then it has been proved that the Onsager's rule can be replaced by the following density quantization rule~\cite{Gao2017a}
\begin{equation}
\rho_\text{semi}(\varepsilon_n)=\left(n+{1\over 2}\right) {eB\over h}\,.
\end{equation}
This suggests that the smooth semiclassical electron density always intersects with the true stepwise electron density at the half-filling points, as shown in Fig.~\ref{fig_ll}.

In general $\rho_\text{semi}$ can be expanded near $B=0$ in power series of $B$.
Up to second order we have
\begin{equation}\label{eq_qr}
\left(n+{1\over 2}\right) {eB\over h}=\frac{S(\varepsilon_n)}{4\pi^2}+B\left.\frac{\partial M}{\partial \mu}\right |_{\mu=\varepsilon_n}+\frac{1}{2}B^2 \left.\frac{\partial \chi}{\partial \mu}\right|_{\mu=\varepsilon_n}\,,
\end{equation}
where $M$ and $\chi$ are the magnetization and magnetic susceptibility respectively, evaluated at zero magnetic field and zero temperature. It is interesting to note that Eq.~\eqref{eq_qr} indicates the nonlinearity in the Landau level fan diagram when plotted as $B$ against $1/n$ due to the appearance of the susceptibility. Such nonlinearity is consistent with experiments~\cite{Taskin2010,Analytis2010,Ren2010,Sacepe2011,Bruene2011,Xiu2011,Xiang2015}.
Based on Eq.~\eqref{eq_qr}, as long as the semiclassical theory can yield the correct susceptibility, it can give correct second order responses in magnetic field.

To evaluate the magnetic susceptibility, we need to calculate the free energy up to second order. In the semiclassical framework, the free energy can be expressed as follows
\begin{align}\label{eq_free}
F&=\int \frac{d\bm k_c}{8\pi^3} \mathcal{D}(g_e(\tilde{\varepsilon})+g_\text{L})\,.
\end{align}
Here the second term $g_\text{L}$ is the Peierls-Landau magnetic free energy:
$g_\text{L}=-(e^2f_0^\prime/48\hbar^2)B_s B_t \epsilon_{s i k}\epsilon_{t j \ell} \alpha_{i j} \alpha_{k\ell}\,,$
where $f_0^\prime$ is the energy derivative of the Fermi distribution function $f_0$. For isotropic bands, the effective mass tensor $\alpha$ is diagonal, and $g_\text{L}$ will reduce to its familiar form~\cite{Mermin1976}. This term originates from higher order corrections to the replacement of the quantum mechanical commutator with the classical Poisson bracket~\cite{Blount1962b}.

After collecting terms at second order in $B$, one can find the following free energy $F=\int g^{\prime\prime} (d\bm k/8\pi^3)$ with
\begin{align}\label{eq_g2}
g^{\prime\prime}&=g_\text{L}+{f_0^\prime\over 2}(\bm B\cdot \bm m)^2-f_0^\prime \bm v_0\cdot \bm P_\text{E}\notag\\
&+f_0\sum_{n\neq 0}{G_{0n}G_{n0}\over \varepsilon_0-\varepsilon_n} +{e^2f_0\over 8m} (B^2 g_{ii}-B_i g_{ij} B_j)\notag\\
&-{3ef_0\over 4\hbar}(\bm B\cdot \bm{\Omega}_0) (\bm B\cdot \bm m)-{e^2f_0\over 8\hbar^2} \epsilon_{s i k} \epsilon_{t j \ell}B_{s} B_{t}g_{ij}\alpha_{k\ell}\,.
\end{align}
The magnetic susceptibility can be readily obtained through its definition
\begin{equation}
\chi_{ij}=-\left.\frac{\partial^2 F}{\partial B_i \partial B_j}\right |_{\mu, T, B=0}\,.
\end{equation}

\begin{figure}[t]
\centering
{\includegraphics[width=\columnwidth]{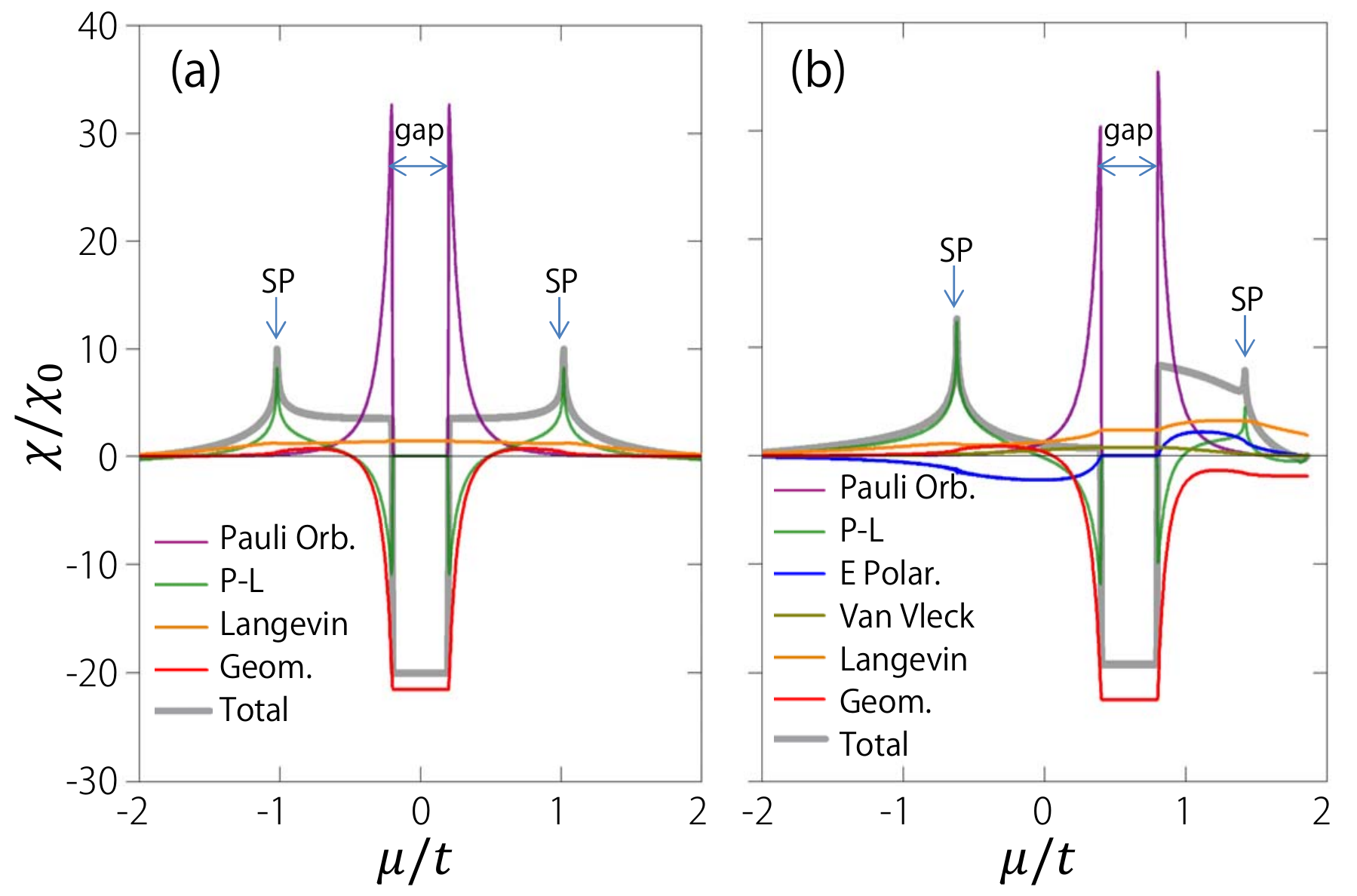}}
\caption{(color online) Orbital magnetic susceptibility for the honeycomb lattice model as a function of $\mu$. $\chi$ is in units of $\chi_0=e^2a^2t/(4\pi^2\hbar^2)$, $a$ is the bond length , $t$ is the nearest neighbour hopping strength, (a) without next nearest neighbour hopping and (b) with next nearest neighbour hopping. Here P-L, E Polar, and SP stand for the Peierls-Landau, energy-polarization, and saddle point, respectively. The susceptibility is consistent with that in Ref.~\cite{Santos2011}. From Ref.~\cite{Gao2015}.}
\label{fig_sus}
\end{figure}

The first two terms in Eq.~\eqref{eq_g2} are Peierls-Landau diamagnetic and Pauli paramagnetic contributions. The physical meaning of the other terms in Eq.~\eqref{eq_g2} can be illustrated by re-expressing it in the Wannier function basis and then taking the atomic insulator limit. In particular, the two terms in the second line reduce to the familiar form of the Van-Vleck paramagnetic free energy and Langevin diamagnetic free energy in atomic physics, respectively. The last term is purely geometrical as it solely comes from the horizontal mixing. The energy polarization and geometrical magnetic free energy approach zero in the atomic limit, indicating that they have no analog in atomic physics and are novel terms in crystals.

The susceptibility obtained from Eq.~\eqref{eq_g2} in the semiclassical framework is consistent with the result derived using other methods. For example, the susceptibility has been calculated in the honeycomb lattice using Hofstadter spectrum or linear response theory~\cite{Santos2011,Raoux2014}. Eq.~\eqref{eq_g2} yields the exactly same result as shown in Fig.~\ref{fig_sus}. In Ref.~\cite{Ogata2015}, it is also proved that Eq.~\eqref{eq_g2} is consistent with the susceptibility from linear response theory~\footnote{In Ref.~\cite{Ogata2015}, it is found that except the energy polarization contribution, all the other terms are consistent. For the energy polarization contribution, Ref.~\cite{Gao2015} contains a typo. When inserting the second order energy in Eq.~\eqref{eq_free}, the energy polarization in Ref.~\cite{Gao2015} has an additional $1/4$ factor by mistake. After removing such factor as given in Eq.~\eqref{eq_g2}, the energy polarization has the same expression with Eq.~(2.31) in Ref.~\cite{Ogata2015}}. This confirms the validity of the semiclassical dynamics under magnetic field.

\subsection{Evaluating positional shift and energy correction in tight-binding models}
The positional shift and energy correction can be implemented in first-principles codes. For this purpose,  besides the energy spectrum, one needs three additional matrix elements: the velocity matrix element $\bm v_{mn}=\langle m|\hat{\bm v}|n\rangle$, the Hessian matrix element $(\Gamma_{ij})_{mn}=\langle u_m|\hat{\Gamma}_{ij}|u_n\rangle$, and the spin Pauli matrix element $\bm \sigma_{mn}=\langle m|\hat{\bm \sigma}|n\rangle$, all of which can be evaluated in first-principles codes in principle~\cite{Marzari1997,Souza2001,Yates2007,Marzari2012}. In the following, we sketch the process of re-expressing the positional shift and energy correction in terms of the energy spectrum and these three matrix elements.

We first examine the positional shift. The interband part of the Berry connection is related to the interband velocity element:
\begin{align}\label{eq_va}
\bm v_{mn}=\frac{i}{\hbar}(\varepsilon_m-\varepsilon_n)\bm {\mathcal{A}}_{mn}\,,\forall \; m\neq n\,.
\end{align}
Using this identity, we manipulate the interband mixing element $G_{n0}$
\begin{align}\label{eq_gn0}
G_{n0}=\frac{ie\bm B\hbar}{2}\cdot\sum_{m\neq 0} \frac{(\bm v_{nm}+\bm v_0\delta_{mn})\times \bm v_{m0}}{\varepsilon_m-\varepsilon_0}+\frac{g\mu_B}{2} \bm B\cdot\bm \sigma_{n0}\,.
\end{align}
Then the first term in the positional shift can be put in the desired form
\begin{equation}\label{eq_ps1}
2\hbar{\rm Im}\sum_{n\neq 0}\frac{\bm v_{0n}G_{n0}}{(\varepsilon_0-\varepsilon_0)^2}\,.
\end{equation}
For the remaining term we note that
\begin{align}\label{eq_ps2}
\epsilon_{ijk}\gamma_{ji\ell}&=\epsilon_{ijk}\partial_ig_{j\ell}\notag\\
&=-2\hbar^3\epsilon_{ijk} \sum_{n\neq 0}{\rm Re}\frac{[(v_i)_0-(v_i)_n](v_j)_{0n}(v_\ell)_{n0}}{(\varepsilon_0-\varepsilon_n)^3}\notag\\
&\quad+\hbar^2\epsilon_{ijk}\sum_{n\neq 0} {\rm Re}\frac{\partial_i(v_j)_{0n}(v_\ell)_{n0}+(j\leftrightarrow \ell)}{(\varepsilon_0-\varepsilon_n)^2}\,.
\end{align}
The only unknown quantity has the form $\partial_i(v_j)_{mn}$, which can be manipulated as follows
\begin{align}\label{eq_dvmn}
\partial_i(v_j)_{mn}=&\sum_{m^\prime\neq m}\frac{\hbar(v_i)_{mm^\prime}(v_j)_{m^\prime n}}{\varepsilon_m-\varepsilon_{m^\prime}}-\sum_{m^\prime\neq n}\frac{\hbar(v_j)_{mm^\prime}(v_i)_{m^\prime n}}{\varepsilon_{m^\prime}-\varepsilon_n}\notag\\
&+\hbar(\Gamma_{ij})_{mn}+i[(\mathcal{A}_i)_m-(\mathcal{A}_i)_n] (v_j)_{mn}\,.
\end{align}
The first three terms are in the desired form. The last term contains the intraband Berry connection which is gauge-dependent. Therefore, it cannot be effectively evaluated in the first-principles codes. However, it can be shown that it does not contribute to Eq.~\eqref{eq_ps2} and hence need not to be evaluated. Equations~\eqref{eq_gn0} to \eqref{eq_dvmn} are enough to transform the positional shift in the desired form.

We now examine the energy correction. Using the following identities
\begin{align}
\bm \Omega_0&=-\hbar^2\sum_{n\neq 0} {\rm Im}\frac{\bm v_{0n}\times \bm v_{n0}}{(\varepsilon_0-\varepsilon_n)^2}\,,\\
\bm m&=-\frac{e\hbar}{2}\sum_{n\neq 0} {\rm Im}\frac{\bm v_{0n}\times \bm v_{n0}}{\varepsilon_0-\varepsilon_n}-\frac{g\mu_B \bm \sigma_{0}}{2}\,,\\
g_{ij}&=\hbar^2\sum_{n\neq 0}{\rm Re}\frac{(v_i)_{0n}(v_j)_{n0}}{(\varepsilon_0-\varepsilon_n)^2}\,,\\
\alpha_{ij}&=(\Gamma_{ij})_{0}+2\sum_{n\neq 0}{\rm Re}\frac{(v_i)_{0n}(v_j)_{n0}}{\varepsilon_0-\varepsilon_n}\,,
\end{align}
together with Eq.~\eqref{eq_va} to \eqref{eq_dvmn}, it is straightforward to put all terms except the energy polarization contribution in Eq.~\eqref{eq_eng2} in the desired form.
We comment that for the tight-binding Hamiltonian, one may have to substitute the last term in Eq.~\eqref{eq_eng2} with Eq.~\eqref{eq_hessian} which has realistic Hessian matrix. In this case, we have
\begin{align}
\varepsilon_\text{H}=&\frac{e^2\hbar^2}{ 8} \sum_{(m,n)\neq 0}  {\rm Re}\frac{(\bm B\times \bm v_{0m})_i (\hat{\Gamma}_{ij})_{mn}(\bm B\times \bm v_{n0})_j}{(\varepsilon_0-\varepsilon_m)(\varepsilon_n-\varepsilon_0)}\notag\\
&+\frac{e^2}{8}\epsilon_{sik}\epsilon_{tj\ell}B_sB_t\sum_{n\neq 0}{\rm Im}[(\Gamma_{ij})_{n0}\partial_\ell (\mathcal{A}_k)_{0n}]\notag\\
&+\frac{e^2}{8}\epsilon_{sik}\epsilon_{tj\ell}B_sB_t\sum_{n\neq 0}{\rm Im}[(\mathcal{A}_k)_{0n} \partial_\ell(\Gamma_{ij})_{n0}]\,.
\end{align}
To evaluate the last two terms, one notes that
\begin{align}
\label{eq_da}\partial_\ell (\mathcal{A}_k)_{0n}&=-\frac{i\hbar^2[(v_\ell)_0-(v_\ell)_n](v_k)_{0n}}{(\varepsilon_0-\varepsilon_n)^2}
+\frac{i\hbar\partial_\ell (v_k)_{0n}}{\varepsilon_0-\varepsilon_n}\,,\\
\label{eq_dgamma}\partial_\ell (\Gamma_{ij})_{n0}&=\sum_{m\neq n}\frac{\hbar(v_\ell)_{nm}(\Gamma_{ij})_{m0}}{\varepsilon_n-\varepsilon_m}-\sum_{m\neq 0}\frac{\hbar(\Gamma_{ij})_{nm}(v_\ell)_{m0}}{\varepsilon_m-\varepsilon_0}\notag\\
&+(\partial_\ell \Gamma_{ij})_{n0}+i[(\mathcal{A}_\ell)_n-(\mathcal{A}_\ell)_0](\Gamma_{ij})_{n0}\,.
\end{align}
Using Eq.~\eqref{eq_da} and Eq.~\eqref{eq_dvmn}, one can put the second term in $\varepsilon_\text{H}$ in the desired form plus extra terms from intraband Berry connection. After inserting Eq.~\eqref{eq_dgamma} into $\varepsilon_\text{H}$, the first term in the second line of Eq.~\eqref{eq_dgamma} does not contribute due to the anti-symmetrization of the spatial indices. The last term in Eq.~\eqref{eq_dgamma} will cancel the extra gauge-dependent terms in Eq.~\eqref{eq_da}. This completes the manipulation of $\varepsilon_\text{H}$.

The manipulation of energy polarization contribution in Eq.~\eqref{eq_eng2} can be done as follows:
\begin{align}
\bm \nabla\cdot \bm P_\text{E}=\frac{1}{2}eB_i\epsilon_{\ell ij}{\rm Re}\sum_{n\neq 0}[\partial_\ell (\mathcal{A}_j)_{0n}G_{n0}+(\mathcal{A}_j)_{0n}\partial_\ell G_{n0}]\,,
\end{align}
where
\begin{align}
\partial_\ell G_{n0}&=-\frac{eB}{2}\sum_{m\neq 0}[\partial_\ell(\bm v_{nm}+\bm v_0\delta_{nm})]\times \bm {\mathcal{A}}_{m0}\notag\\
&-\frac{eB}{2}\sum_{m\neq 0}(\bm v_{nm}+\bm v_0\delta_{nm})\times (\partial_\ell\bm {\mathcal{A}}_{m0})\notag\\
&+\frac{g\mu_B \hbar\bm B}{2}\cdot \left[\sum_{m\neq n}\frac{(v_\ell)_{nm}\bm \sigma_{m0}}{\varepsilon_n-\varepsilon_m}
-\sum_{m\neq 0}\frac{(\bm \sigma)_{nm}(v_\ell)_{m0}}{\varepsilon_m-\varepsilon_0}\right]\notag\\
&+\frac{g\mu_B \hbar\bm B\cdot\bm \sigma_{n0}}{2} i[(\mathcal{A}_\ell)_n-(\mathcal{A}_\ell)_0]\,.
\end{align}
Again, terms contain intraband Berry connection can be ignored as they will cancel each other in the end.

We comment that if one starts from the Schr$\ddot{\rm o}$dinger Hamiltonian, the Hessian matrix can be directly evaluated without using wave functions in first-principles codes. For Schrodinger Hamiltonian, $\hat{\Gamma}_{ij}=1/m\delta_{ij}$. Therefore, one has $(\Gamma_{ij})_{mn}=1/m \delta_{ij}\delta_{mn}$. Then for the semiclassical dynamics, one only needs to evaluate the velocity and spin matrix elements using output from first-principles codes.

\begin{figure}
  \centering
  \includegraphics[width=\columnwidth]{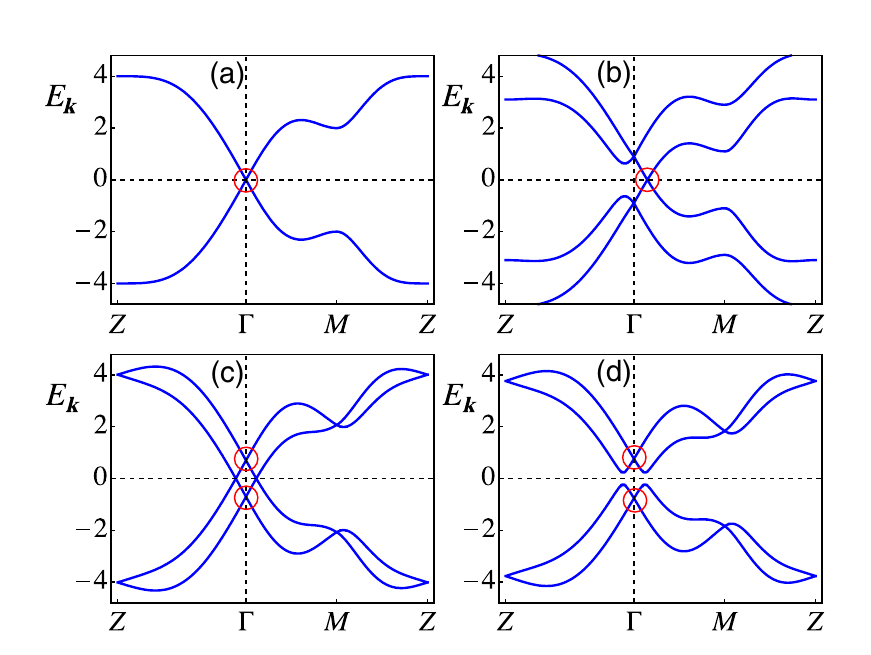}\\
  \caption{Band structure for (a) Dirac semimetal, (b) Weyl semimetal by breaking time reversal symmetry, (c) Weyl semimetal by breaking inversion symmetry, and (d) semiconductor. From Ref.~\cite{Vazifeh2013}.}\label{fig_eng0}
\end{figure}

As a concrete example, we consider the following tight-binding Hamiltonian describing a 3D TI in the Bi$_2$Se$_3$ family on a simple cubic lattice~\cite{Fu2010,Qi2011,Vazifeh2013}
\begin{align}\label{eq_h0}
\hat{H}_\text{tb}=2\lambda \sigma_z(s_x\sin k_y-s_y \sin k_x)+2\lambda_z \sigma_y \sin k_z+\sigma_x M_k\,,
\end{align}
where $\bm \sigma$ and $\bm s$ are Pauli matrices in orbital and spin space respectively, and $M_k=\epsilon-2t(\cos k_x+\cos k_y +\cos k_z)$. At $\epsilon=6t$, the above Hamiltonian is at the phase transition point between trivial phase and topological insulator phase, and supports a Dirac point at the $\Gamma$ point. The Dirac point can split into Weyl point if the following additional term is added to the Hamiltonian
\begin{equation}\label{eq_h1}
\hat{H}_\text{break}=b_0 \sigma_ys_z+b_zs_z\,.
\end{equation}
These two terms break inversion and time reversal symmetry, respectively. The resulting Weyl semimetal phase is displayed in Fig.~\ref{fig_eng0}.

\begin{figure}[t]
  \centering
  \includegraphics[width=\columnwidth]{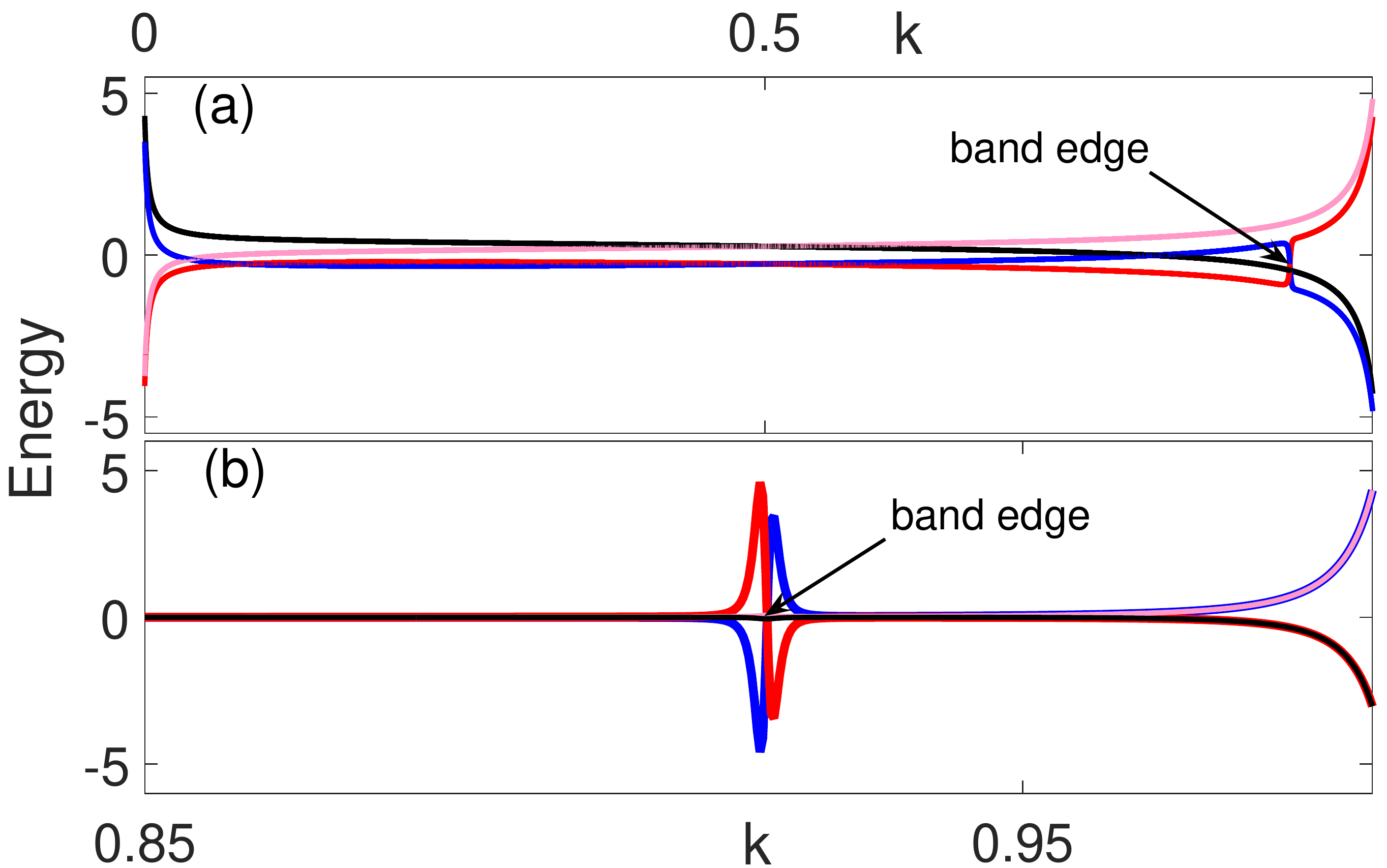}\\
  \caption{First order (Panel a) and second order (Panel b) correction to the band energy. Parameters are chosen as follows: $\lambda=\lambda_z=1$, $t=0.5$, $\epsilon=2.8$, $b_z=0$, $b_0=0.8$, the flux of $B$ through the unit cell $\phi=eBa^2/\hbar=3.79\times 10^{-4}$, the spin Zeeman energy is taken to be $g\mu_BB=3.86\times 10^{-4}\lambda$ which is roughly at the same order of the orbital Zeeman energy. For Panel~(a), energy is in units of $10^{-3}\lambda$ and for Panel~(b) energy is in units of $10^{-4}\lambda$. The $x$-axis is along $Z$-$\Gamma$ direction, similar with Fig.~\ref{fig_eng0}. Here we take the total length of $Z\Gamma$ to be unity and record the relative position along $Z\Gamma$ direction. In Panel (a), the energy correction is plotted along all the $Z\Gamma$ line while in Panel (b), only a portion of $Z\Gamma$ line near the band gap and band crossing points are plotted, to better illustrate the structure of the energy correction. The corrections for lowest to highest bands are represented in black, blue, red, pink colors, respectively. Near the $\Gamma$ point, black and red curves coincide, and so do the blue and pink curves.}\label{fig_eng12}
\end{figure}

Here we choose the semiconductor phase and calculate the first and second order correction to the band energy, as shown in Fig.~\ref{fig_eng12}. It can be seen that both first and second order correction vary drastically near the band crossing points and the small band gap. In fact, both corrections diverge at the $\Gamma$ point due to the band crossing. Across the small global band gap as shown in Fig.~\ref{fig_eng0}, both first and second order corrections change sign. It is also interesting to note that near the $\Gamma$ point, the lowest two bands have similar first order corrections, but are opposite to those of the remaining two bands. This property is consistent with the calculation of magnetic moment for the low-energy Dirac model. In comparison, the second order correction have different properties. The lowest and the second highest band have similar second order corrections, opposite to those of the remaining two bands. This can be most easily seen from Eq.~\eqref{eq_eng2}. The Berry curvature, effective mass tensor and energy gap reverse sign for the two bands forming a Dirac cone, leading to the sign change in their second order energy correction.

\section{Nonlinear charge current}
With the semiclassical theory up to second order at hand, we can derive the nonlinear charge current up to third order, i.e. conductivity up to second order in fields. We focus on the semiclassical regime with weak electromagnetic fields. There is one important issue that we want to discuss before the detailed derivation of nonlinear currents. In the previous section, the semiclassical theory is generally derived for a single Bloch band that is well separated from all the other bands. In reality, this condition is hardly met as band-crossing points are generally present. In this case, the previous semiclassical theory can still be used to derive the conductivity as long as the Fermi surface is not close to those band-crossing points. The reason is as follows. As a perturbation theory, the semiclassical theory may fail when the band gap is small, such as near the band crossing points. However, the conductivity is a Fermi surface property. As a result, as long as the band-crossing points are deep inside the Fermi sea, the perturbation theory still works near the Fermi surface.

In this section, we will first establish the general theory of the charge current beyond the linear order. Then we will discuss several important examples, including the response of the anomalous Hall conductivity to electromagnetic fields, linear magnetoresistance in time-reversal-broken materials, chiral anomaly in Weyl semimetals and two different mechanisms for the negative longitudinal magnetoresistance: the intrinsic quadratic magnetoresistance and the current jetting.

\subsection{General theory of nonlinear charge currents}
The current in the semiclassical theory reads
\begin{equation}\label{eq_cur}
\bm J=-e\int \frac{d\bm k}{8\pi^3} \mathcal{D}\dot{\bm r} f \,,
\end{equation}
where $f$ is the electron distribution function, which is the Fermi function in equilibrium. Here and hereafter, we drop the subscript $c$ in the center of mass position $\bm r_c$ and momentum $\bm k_c$ for simplicity.
From the semiclassical dynamics in Eq.~\eqref{eq_dotrc2} and \eqref{eq_dotkc2}, one can show that
\begin{equation}\label{eq_dr}
\mathcal{D}\dot{\bm r}=\tilde{\bm v}+e\bm E\times {\bm \Omega}^\text{t}+\frac{e}{\hbar}\left(\tilde{\bm v}\cdot {\bm \Omega}^\text{t}\right)\bm B\,,
\end{equation}
where $\tilde{\bm v}=\partial \tilde{\varepsilon}/\hbar\partial \bm k$ is the modified band velocity. Here and hereafter, we ignore the band index $0$ in relevant intraband quantities for simplicity and the subscript $0$ only has the meaning of zeroth order.

The remaining factor in the current is the distribution function $f$, which is typically solved from the Boltzmann equation. If the sample is homogeneous and reaches a steady state under external fields, under the relaxation time approximation, the Boltzmann equation reads
\begin{equation}
\dot{\bm k}\cdot \frac{\partial f}{\partial \bm k}= \left. \frac{df}{dt}\right |_\text{collision}=-\frac{f-f_0}{\tau}\,.
\end{equation}
Here the argument of equilibrium distribution $f_0$ is the modified band energy $\tilde{\varepsilon}$. Solving the Boltzmann equation perturbatively, one has~\cite{Pal2010}
\begin{equation}\label{eq_dis}
f=\sum_{m=0}^\infty (-\tau \dot{\bm k}\cdot \bm \partial_{\bm k})^m f_0(\tilde{\varepsilon})\,.
\end{equation}

Equations~\eqref{eq_cur}, \eqref{eq_dr} and \eqref{eq_dis} are enough to derive the current except the anomalous Hall current up to third order of the external fields, i.e. the conductivity up to second order. Other than the anomalous Hall current, the equilibrium distribution $f_0$ alone does not contribute to the current. Therefore, at least the first order correction to $f_0$ is required, which automatically contains an external field due to $\dot{\bm k}$. As a result, one only needs to evaluate the factor $\mathcal{D}\dot{\bm r}$ up to second order, which is given in the semiclassical dynamics. For the anomalous Hall conductivity, the semiclassical theory can only yields its first order correction, as only the first order correction to Berry curvature is obtained in the semiclassical theory.

To explicitly write out the solution in Eq.~\eqref{eq_dis} up to third order in fields, we note that
\begin{align}
-\tau\dot{\bm k}&=\frac{\tau}{\hbar} \frac{e\bm E+e\tilde{\bm v}\times \bm B+\frac{e^2}{\hbar}(\bm E\cdot \bm B)\bm \Omega^\text{t}}{1+\frac{e}{\hbar}\bm B\cdot \bm \Omega^\text{t}}\,.
\end{align}
Therefore, we write the distribution function as follows
\begin{equation}
f=f_0(\tilde{\varepsilon})+f_1+f_2+f_3\,,
\end{equation}
where the subscript $1$, $2$, and $3$ stand for different orders in $\tau$, and
\begin{widetext}
\begin{align}
\label{eq_f1}f_1=&\frac{\tau}{\hbar} \frac{e\bm E+\frac{e^2}{\hbar}(\bm E\cdot \bm B)\bm \Omega^\text{t}}{1+\frac{e}{\hbar}\bm B\cdot \bm \Omega^\text{t}}\cdot \bm \partial f_0(\tilde{\varepsilon})\,,
\end{align}
\begin{align}
\label{eq_f2}f_2=&\frac{e^2\tau^2 [E_i+(\bm v\times \bm B)_i]E_j}{\hbar^2}\partial_i\partial_j f_0(\varepsilon-\bm B\cdot \bm m)\notag\\
&+\frac{2e^2\tau^2 [E_j+(\bm v\times \bm B)_j]\partial_i\partial_j f_0(\varepsilon)}{\hbar^2}\left[\frac{E_i+(\bm v\times \bm B)_i}{1+\frac{e}{\hbar}\bm B\cdot \bm \Omega}+\frac{e}{\hbar}(\bm E\cdot \bm B)\Omega_i\right]\notag\\
&+\frac{e^2\tau^2 [E_j+(\bm v\times \bm B)_j]\partial_i f_0(\varepsilon)}{\hbar^2}\partial_j\left[\frac{E_i+(\bm v\times \bm B)_i}{1+\frac{e}{\hbar}\bm B\cdot \bm \Omega}+\frac{e}{\hbar}(\bm E\cdot \bm B)\Omega_i\right]\,,\\
\label{eq_f3}f_3=&\frac{e^3\tau^3E_i}{\hbar^3}[(\bm E+\bm v\times \bm B)\cdot \bm \partial]^2 \partial_{i}f_0(\varepsilon)\,.
\end{align}
\end{widetext}

There is another contribution to the nonlinear conductivity. The system can reach a thermodynamic equilibrium together with a uniform magnetic field. During such process, if the sample remains charge neutral, the carrier density is fixed. As the magnetic field modifies the band energy, the chemical potential has to vary accordingly.

Without loss of generality, we assume $\bm B$ is along the $z$ direction, i.e. $B=B_z$. At first order in $B$, the change in chemical potential can be fixed using the following expansion of the electron density $n$
\begin{equation}\label{eq_nexp1}
dn=\left.\frac{\partial n}{\partial B}\right |_{\mu,T,B=0} B+\left.\frac{\partial n}{\partial \mu}\right |_{T,B=0}  \mu_{(1)}=0\,.
\end{equation}
Note that $\left.\frac{\partial n}{\partial \mu}\right |_{T,B=0}=g(\mu)$, which is simply the density of states at the chemical potential for the unperturbed band. Moreover, $\left.\frac{\partial n}{\partial B}\right |_{\mu,T}$ can be connected to the magnetization through the Maxwell equation:
\begin{equation}\label{eq_dndb1}
\left.\frac{\partial n}{\partial B}\right |_{\mu,T}=\left. \frac{\partial M_z}{\partial \mu} \right |_{T,B}\,.
\end{equation}
We comment that for insulators at zero temperature, $-\frac{\partial M_z}{\partial \mu}\rightarrow \sigma_{xy}$, and Eq.~\eqref{eq_dndb1} reduces to the familiar Streda formula.
Combining Eq.~\eqref{eq_nexp1} and \eqref{eq_dndb1}, we obtain
\begin{equation}\label{eq_mu1}
\mu_{(1)}=-\frac{(\partial M_z/\partial \mu)_{T,B=0}}{g(\mu)}\,.
\end{equation}

At second order in $B$, the expansion of $n$ reads
\begin{align}
dn&=\left.\frac{\partial n}{\partial \mu}\right |_{T,B=0}\mu_{(2)}+\frac{1}{2} \left.\frac{\partial^2n}{\partial B^2} \right |_{T,\mu,B=0} B^2\notag\\
&+\frac{1}{2}\left.\frac{\partial^2 n}{\partial \mu^2}\right |_{T,B=0}[\mu_{(1)}]^2+\left.\frac{\partial^2n}{\partial B\partial \mu}\right |_{T,B=0} B\mu_{(1)}\,.
\end{align}
$\mu_{(2)}$ can be solved by setting the above equation equal to zero. The result can be simplified using Eq.~\eqref{eq_dndb1} and the definition of the isothermal magnetic susceptibility $\chi_{zz}=\partial M_z/\partial B$, yielding
\begin{equation}\label{eq_mu2}
\mu_{(2)}=-\frac{1}{2}\frac{B^2\frac{\partial \chi_{zz}}{\partial \mu}+\frac{\partial g}{\partial \mu}[\mu_{(1)}]^2+2\frac{\partial^2 M_z}{\partial \mu^2} B\mu_{(1)}}{g(\mu)}\,.
\end{equation}

These change in chemical potential will also contribute to the nonlinear conductivity:
\begin{align}\label{eq_jm}
\bm J_\mu=&-e\mu_{(1)} \frac{\partial}{\partial\mu} \int \frac{d\bm k}{8\pi^3}\mathcal{D}\dot{\bm r}f\notag\\
&+\frac{e^2\tau}{\hbar} \int \frac{d\bm k}{8\pi^3} \bm v (\bm E\cdot \bm v)\left\{\mu_{(2)}f_0^{\prime\prime}-\frac{1}{2}[\mu_{(1)}]^2 f_0^{\prime\prime\prime}\right\}\,,
\end{align}
where $f_0^{\prime\prime}$ and $f_0^{\prime\prime\prime}$ are second and third order energy derivatives of $f_0$.

Equation~\eqref{eq_cur} with Eq.~\eqref{eq_dr} and \eqref{eq_f1}-\eqref{eq_f3}, combined with Eq.~\eqref{eq_mu1}, \eqref{eq_mu2}, and \eqref{eq_jm} yields a complete description of the nonlinear conductivity up to second order in the framework of the semiclassical theory. This procedure gives a complete account for the drifting part in the Boltzmann equation. However, it ignores the dependence of the collision integral on electromagnetic fields. Nevertheless, the semiclassical theory can yield a qualitatively valid result and offers a fresh perspective of how the band properties beyond the spectrum affects the conductivity.

Finally, we comment that to get the corresponding change in the resistivity, one can invert the conductivity tensor. For example, if change in conductivity $\delta \sigma_{xx}$ is obtained, the corresponding change in resistivity is
\begin{equation}
\frac{\delta \rho_{xx}}{\rho_{xx}}=-\frac{\delta \sigma_{xx}}{\sigma_{xx}}\,,
\end{equation}
provided that the Hall conductivity $\sigma_{xy}$ is much smaller than the longitudinal conductivity $\sigma_{xx}$.

\subsection{Nonlinear anomalous Hall conductivity}
The anomalous Hall effect refers to a Hall-type current in ferromagnets solely driven by an electric field. It is a topic under extensive studies~\cite{Nagaosa2010}. In the past, there are three mechanisms identified: the intrinsic contribution, the skew-scattering contribution, and the side-jump contribution. Here we focus on the intrinsic contribution, which is due to the anomalous velocity in Eq.~\eqref{eq_dotrc}~\cite{Xiao2010}. From the perspective of the semiclassical theory, the Berry curvature acts as the magnetic field in the momentum space and bends the electron trajectory likewise, leading to a Hall-type current. The anomalous Hall effect has also been studied in noncolinear antiferromagnets, in which the net spin magnetization vanishes but the orbital magnetization is still present~\cite{Ohgushi2000,Shindou2001,Taillefumier2006,Kalitsov2009,Takatsu2010,Udagawa2013,Chen2014,Suzuki2017,Guo2017}.

From symmetry consideration, the anomalous Hall effect requires broken time-reversal symmetry. In other words, it requires either a net spin magnetization as in ferromagnets or orbital magnetization as in noncolinear antiferromagnets. It is forbidden if time reversal symmetry is present. However, the nonlinear Hall effect may still be present, due to the manipulation of the magnetization through electromagnetic fields.

In systems with both time reversal and inversion symmetry, a net magnetization can be induced through the magnetic susceptibility by a magnetic field. This magnetization can then lead to a Hall-type current. The analytical expression for this correction can be obtained by plugging the anomalous velocity (second term in Eq.~\eqref{eq_dr}) into the current in Eq.~\eqref{eq_cur}. The result reads~\cite{Gao2014}
\begin{equation}\label{eq_aheb}
\bm J=\frac{e^2}{\hbar}\bm E\times \int [\hbar\bm v\times \bm {\mathcal{A}}^\prime(\bm B)+\bm \Omega(\bm B\cdot \bm m)] f_0^\prime \frac{d\bm k}{8\pi^3}\,,
\end{equation}
where $\bm {\mathcal{A}}^\prime(\bm B)$ stands for the part of $\bm {\mathcal{A}}^\prime$ solely dependent on $\bm B$ and $\bm m$ is the spin plus orbital magnetic moment. It can be explicitly checked that compared with the susceptibility in Eq.~\eqref{eq_g2}, the second term in Eq.~\eqref{eq_aheb} is part of the geometric contribution to the magnetic susceptibility. Moreover, the first term is also part of the Van-Vleck and energy polarization contributions to the susceptibility. This current yields the conductivity $\sigma_{(ij,k),0}$ in Eq.~\eqref{eq_jeb}.

This current has exactly the same dependence on electromagnetic fields with the ordinary Hall current. However, they have different origins. The ordinary Hall current is due to the Lorentz force and has the form of $\omega_c \tau$ with $\omega_c$ being the cyclotron frequency. In contrast, the current in Eq.~\eqref{eq_aheb} requires a nontrivial structure in the momentum space and does not involve the relaxation time. In fact, the ratio of the resistivity $\rho_{xy}^\prime$ from Eq.~\eqref{eq_aheb} and the ordinary Hall resistivity $\rho_{xy}^\text{ord}$ can be put in the following form
\begin{align}\label{eq_scale}
\frac{\rho^\prime_{xy}}{\rho_{xy}^\text{ord}}=\left(\rho_{xx} \frac{e^2}{4h}\right)^2 S(\mu)\,,
\end{align}
where the first factor is universal and $S(\mu)$ is a model-dependent factor but independent of the scattering process. The universal scaling factor means that $\rho^\prime_{xy}$ will dominate the ordinary Hall effect when $\rho_{xx}$ is large, i.e. in dirty metals/semiconductors with relatively small relaxation time. In a typical Hall-bar measurement set-up, both $\rho_{xy}^\prime$ and $\rho_{xy}^\text{ord}$ will contribute to the Hall current. Therefore, the total Hall resistivity should be dependent on the relaxation time. To differentiate one contribution from the other, one should change the universal scaling factor through temperature, film thickness, or doping, and measure the scaling behaviour based on Eq.~\eqref{eq_scale}.

\begin{figure}
  \centering
  \includegraphics[width=\columnwidth]{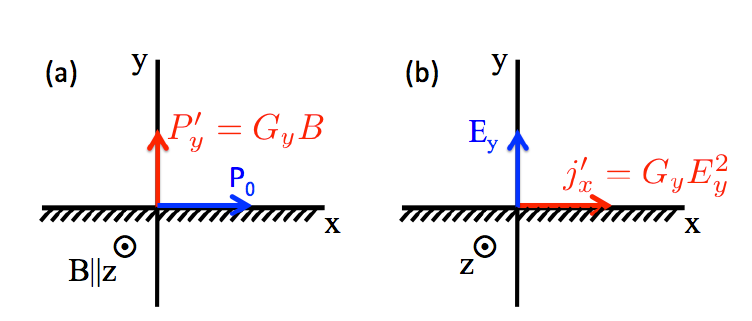}\\
  \caption{Magnetoelectric effect~(Panel a) and the related nonlinear anomalous Hall induced by the electric field~(Panel b). Originally the magnetoelectric effect is shown as the polarization induced by a magnetic field, which is the same as a magnetization along $\hat{z}$ induced by an electric field along $\hat{y}$ through the magnetoelectric coefficient $G_{y}$. From Ref.~\cite{Gao2014}.}\label{fig_nahe}
\end{figure}

In systems that simultaneously break the time reversal and inversion symmetry but preserve the combined symmetry, the anomalous Hall current also vanishes. This symmetry is exactly the magnetoelectric symmetry which allows a magnetization to be induced by an electric field through the magnetoelectric coefficient~\cite{Landau1984}. This magnetization can then lead to a Hall-type current. The analytical expression can be obtained in a similar way with the magnetic-field-correction to anomalous Hall effect. The result reads~\cite{Gao2014}
\begin{equation}\label{eq_nahe}
\bm J=e^2\bm E\times \int [\bm v\times \bm {\mathcal{A}}^\prime(\bm E)] f_0^\prime \frac{d\bm k}{8\pi^3}\,,
\end{equation}
where $\bm {\mathcal{A}}^\prime(\bm E)$ stands for the part of $\bm {\mathcal{A}}^\prime$ solely dependent on $\bm E$. This current yields the conductivity $\sigma_{ijk,0}$ in Eq.~\eqref{eq_jee}. The connection between this Hall-type current and the magnetoelectric effect can be mostly seen for a two-band model, as shown in Fig.~\ref{fig_nahe}.

In noncentrosymmetric but time-reversal-invariant materials, a net magnetization can be induced in the nonequilibrium steady state through the Edelstein effect. Such magnetization can further lead to a Hall-type current. The analytical expression can be obtained by plugging Eq.~\eqref{eq_dr} and \eqref{eq_f1} into Eq.~\eqref{eq_cur} and keeping terms at second order in $\bm E$, and reads~\cite{Inti2015}
\begin{equation}\label{eq_eahe}
\bm J=\frac{e^3}{\hbar}\tau\bm E\times \int \bm \Omega (\bm E\cdot \bm v) f_0^\prime \frac{d\bm k}{8\pi^3}\,.
\end{equation}
This current yields the conductivity $\sigma_{ijk,1}$ in Eq.~\eqref{eq_jee}. Interestingly, the requirement of the mirror symmetry in Fig.~\ref{fig_nahe}(b) also works for this current. But one should keep in mind that for the current in Eq.~\eqref{eq_nahe}, as the time reversal symmetry is broken, the inherent spin texture in the sample is also subjected to the mirror operation. The observation of this nonlinear Hall current has been reported in several recent experiments~\cite{Xu2018,Kang2018,Ma2018}. We comment that in Eq.~\eqref{eq_eahe}, $\bm v\bm \Omega$ together constitutes a pseudotensor which represents the first order moment of the Berry curvature in the momentum space. Hence it is referred to as the Berry curvature dipole. This Berry curvature dipole, together with the magnetic moment dipole also play essential roles in the study of the natural optical activity~\cite{Landau1984,Malashevich2010,Orenstein2013,Ma2015,Zhong2015,Zhong2016}.

Finally, using the quantum kinetic equations for the evolution of the density matrix under both electric field and disorders, it has been found that in noncentrosymmetric materials, besides the Berry curvature dipole contribution, there are also nonlinear side jump and skew scattering mechanisms responsible for the nonlinear anomalous Hall effect at second order of electric field~\cite{Nandy2019}. It can compete with the nonlinear anomalous Hall effect due to Berry curvature dipole.

\subsection{Linear Magnetoresistance}
Linear magnetoresistance, referring to the linear dependence of the resistance on magnetic field, was first observed in simple metals~\cite{Reitz1967,Penz1968,Jones1969} and under continuous studies afterwards. There is a surge of interest in recent years, as it appears in various novel materials, such as multilayer graphene~\cite{Friedman2010}, topological insulators~\cite{Qu2010,Tang2011,Wang2012,Tian2014}, and Weyl/Dirac semimetals~\cite{He2014,Liang2014,Feng2015,Novak2015,Narayanan2015,Huang2015}. It has been shown that the linear magnetoresistance can arise in the ultra-quantum regime for Dirac systems when only the lowest Landau level is partially filled~\cite{Abrikosov1988,Abrikosov2000,Wang2012a}, as well as in inhomogeneous samples with mobility fluctuation~\cite{Herring1960,Parish2003,Porter2012,Kozlova2012}. In the semiclassical regime, linear magnetoresistance is subject to a symmetry requirement. Generally speaking Onsager relation requires that $\sigma_{xx}(\bm B)=\sigma_{xx}(-\bm B)$, which forbids odd power dependence of the conductivity on magnetic field. However, if the time reversal symmetry is broken and a magnetization is present, Onsager relation becomes $\sigma_{xx}(\bm B,\bm M)=\sigma_{xx}(-\bm B,-\bm M)$, which can be satisfied in principle for terms with odd powers of $B$.

To derive the linear magnetoresistance, or equivalently the linear magnetoconductivity in the semiclassical regime, one collects terms linear in $B$ in the current in Eq.~\eqref{eq_cur}. To simplify the result, we first consider the transverse magnetoresistance and assume that the electric field is along $\hat{x}$ direction and $\bm B$ is along $\hat{z}$ direction. This configuration will yield the following conductivity (here we use the definition in Eq.~\eqref{eq_jeb})~\cite{Chen2015}
\begin{align}
\sigma_{(xx,z),1}&=\frac{e^2}{\hbar}\int\frac{d\bm k}{8\pi^3}\left[2v_x\frac{\partial m_z}{\partial k_x}+\frac{e}{\hbar}(v_x)^2\Omega_z\right]f_0^\prime\notag\\
&+\frac{e^2}{\hbar}\int\frac{d\bm k}{8\pi^3}(v_x)^2 (m_z+\delta \mu_{(1)}/B_z) f_0^{\prime\prime}\,.\\
\sigma_{(xx,z),2}&=-e^3 \int\frac{d\bm k}{8\pi^3}(v_xv_y\alpha_{xx}-v_x^2\alpha_{xy}) f_0^\prime=0\,.
\end{align}

For longitudinal magnetoresistance, we assume that the electric field is parallel to the magnetic field, i.e. it is along $\hat{z}$ direction. Then the magnetoconductivity reads
\begin{align}
\sigma_{(zz,z),1}&=\frac{e^2}{\hbar}\int\frac{ d\bm k}{8\pi^3}\left[2v_z\frac{\partial m_z}{\partial k_z}+\frac{e}{\hbar}v_z(v_z\Omega_z-2\bm v\cdot \bm \Omega)\right]f_0^\prime\notag\\
&+\frac{e^2}{\hbar}\int\frac{d\bm k}{8\pi^3}(v_z)^2 (m_z+\delta \mu_{(1)}/B_z) f_0^{\prime\prime}\,.\\
\sigma_{(zz,z),2}&=-e^3 \int\frac{d\bm k}{8\pi^3}(v_zv_y\alpha_{xz}-v_zv_x\alpha_{yz}) f_0^\prime=0\,.
\end{align}

Both $\sigma_{(xx,z),1}$ and $\sigma_{(zz,z),1}$ are clearly violations to the Kohler's rule~\cite{Pippard1989}. In other words, the magnetic field affects the conductivity not through the Lorentz force, but through the two extra corrections in the first order semiclassical equations of motion. On one hand, it couples to the $k$-dependent magnetic moment to modifies the band energy and hence band velocity. On the other hand, it couples to the Berry curvature and changes the phase of space density of states, or equivalently, the carrier density. In the longitudinal magnetoconductivity, there is an additional term with the factor $\bm v\cdot \bm \Omega$. This actually measures the flux strength of the momentum space magnetic field $\bm \Omega$. This unique contribution to the longitudinal magnetoresistance is due to the coupling between the real and momentum space lorentz force, i.e. the magnetic field along $\hat{z}$ can first bend the velocity along $\hat{x}$ (or $\hat{y}$) to the direction $\hat{y}$ (or $\hat{x}$), which is further bent to the $\hat{z}$ direction due to the momentum space magnetic field $\Omega_x$ (or $\Omega_y$). Finally, we comment that both $\sigma_{(xx,z),1}$ and $\sigma_{(zz,z),1}$ are linear in magnetic moment and Berry curvature, which are odd in time reversal operation. As a result, these two conductivities only appear in materials that break time reversal symmetry.

The fact that $\sigma_{(xx,z),2}$ and $\sigma_{(zz,z),2}$ vanish identically is consistent with the Onsager's relation, as formally they do not need to break the time reversal symmetry according to Table.~1. We comment that if a detailed consideration of scattering process is performed, there is a contribution to $\sigma_{(xx,z)}$ that is of second order in $\tau$, due to the skew scattering process~\cite{Chen2015}. Therefore, it requires a nontrivial antisymmetric part of the scattering probability $W_{\bm k\bm k^\prime}$ and hence the breaking of time reversal symmetry.

\begin{figure}
  \centering
  \includegraphics[width=0.9\columnwidth]{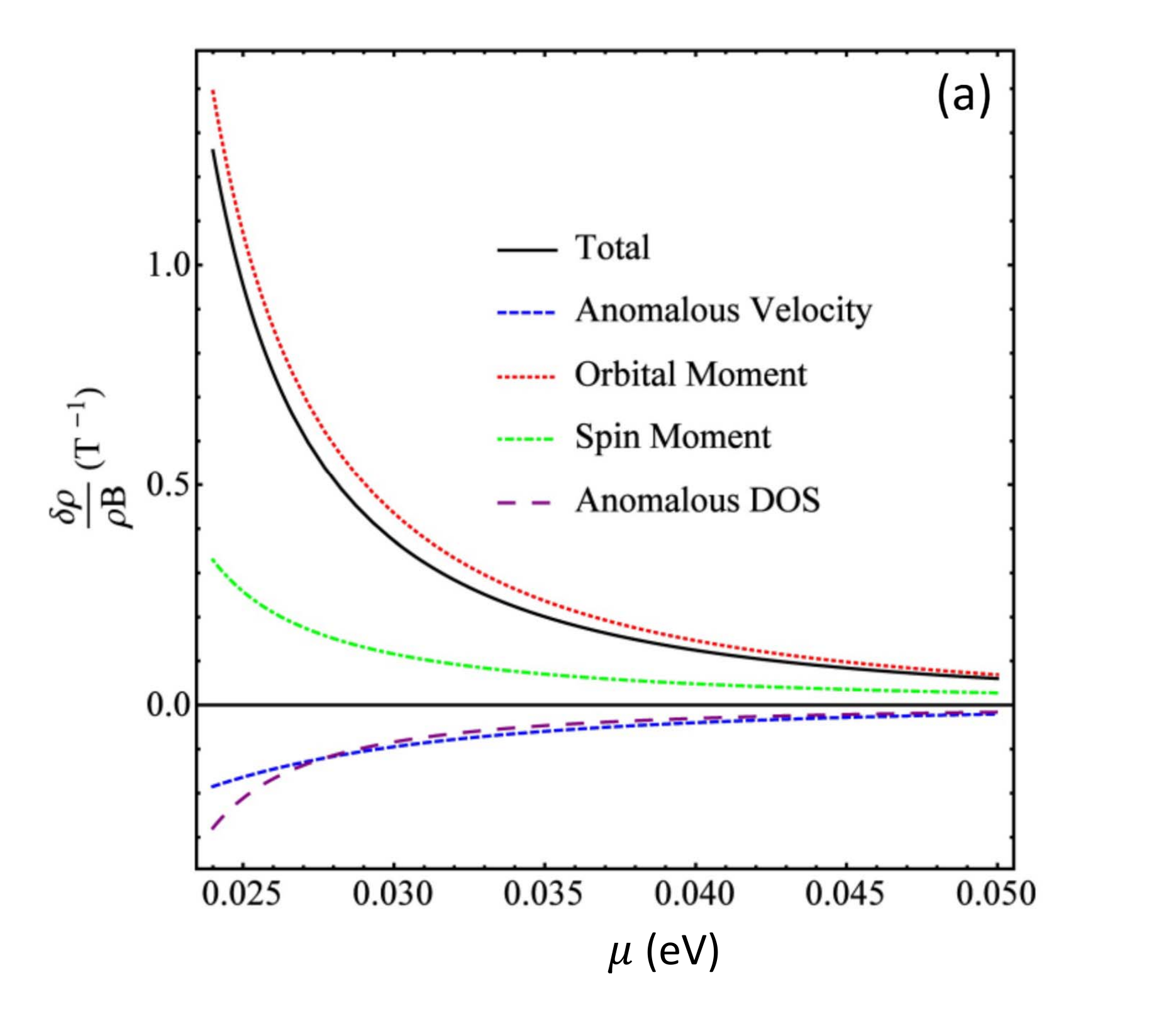}\\
  \caption{Linear magnetoresistance from $\sigma_{(xx,z),1}$ as a function of chemical potential. It is calculated for the surface state of the ferromagnetic topological insulator with the Hamiltonian $\hat{H}=\hbar v_\text{F} (\sigma_xk_x+\sigma_yk_y)+\Delta \sigma_z$. $\hbar v_\text{F}=4.1{\rm eV}\cdot {\rm A}$. $\Delta=20{\rm meV}$. From Ref.~\cite{Chen2015}.}\label{fig_lmr}
\end{figure}

When the chemical potential approaches the hot spot of the Berry curvature and the magnetic moment, the latter can have a large contribution to the linear magnetoresistance, as shown in Fig.~\ref{fig_lmr}. As the chemical potential approaches the band edge, the Berry curvature and magnetic moment increases and so does the linear magnetoresistance. However, the semiclassical theory is only valid when $(\mu-\Delta)\tau/\hbar>1$. Otherwise, the scattering from different impurities will be correlated and a fully quantum mechanical treatment should be used.

\subsection{Negative longitudinal magnetoresistance induced by chiral anomaly}
It is well known that the chemical potential difference in real space can lead to a current. Its counterpart in momentum space, i.e. the current due to the chemical potential difference in momentum space, has completely different origins and attracts great attention recently as it manifests in topological semimetals.

To understand this effect, we note that from the semiclassical theory, the Berry curvature is the analog of the magnetic field in the momentum space. The flux of the Berry curvature,
defined as $(1/2\pi)\oint d\bm S\cdot \bm \Omega$, can be nonzero through a closed surface, indicating the appearance of a monopole charge in the momentum space. This is most easily realized by a pair of Weyl points
\begin{align}\label{eq_dirac}
\hat{H}=\lambda_i v\bm k\cdot \bm \sigma\,.
\end{align}
where $\lambda_i=\pm 1$ stands for a positive or negative charge, or equivalently, the chirality of the Weyl node. For such pair of Weyl nodes, there is a current driven by the magnetic field. This is the chiral magnetic current, derived using the last term in Eq.~\eqref{eq_dr}
\begin{align}\label{eq_cme}
J_\text{CME}=-\frac{e^2}{\hbar}\int \frac{d\bm k}{8\pi^3} (\bm v\cdot \bm \Omega)\bm Bf_0=-\frac{e^2\bm B}{4\pi^2\hbar^2}\sum_i \lambda_i \mu_i\,,
\end{align}
where $\mu_i$ is the chemical potential for each Weyl node. This current has to vanish in equilibrium~\cite{Vazifeh2013}, even when the two Weyl nodes are shifted in energy and the relative chemical potential is different.

However, away from equilibrium, the chemical potential for different valleys can respond to external fields in different ways, leading to an inhomogeneous chemical potential distribution in the momentum space and hence a net current through $J_\text{CME}$~\cite{Son2013}. To see this, we first write down the Boltzmann equation:
\begin{equation}\label{eq_ca}
\dot{\bm k}\cdot \bm \partial f_0=-\int d\bm k^\prime \mathcal{D}(\bm k^\prime) W_{\bm k\bm k^\prime}[f(\bm k)-f(\bm k^\prime)]\,,
\end{equation}
where $W_{\bm k\bm k^\prime}$ is the scattering probability.
Using the semiclassical equations of motion in Eq.~\eqref{eq_dotrc} and \eqref{eq_dotkc}, one finds that
\begin{equation}\label{eq_dk}
\mathcal{D}\dot{\bm k}=-\frac{e}{\hbar}\bm E-\frac{e}{\hbar}\tilde{\bm v}\times \bm B-\frac{e^2}{\hbar^2}(\bm E\cdot \bm B)\bm \Omega\,.
\end{equation}
Therefore, the Boltzmann equation becomes
\begin{align}\label{eq_bz}
&\frac{e}{\mathcal{D}(\bm k)\hbar}\left[ \bm E+\frac{e}{\hbar}(\bm E\cdot \bm B)\bm \Omega\right]\cdot \bm \partial f_0\notag\\
=&\int d\bm k^\prime \mathcal{D}(\bm k^\prime) W_{\bm k\bm k^\prime}[f(\bm k)-f(\bm k^\prime)]\,.
\end{align}

In multivalley systems such as a pair of Weyl nodes, a special solution to the above equation may emerge. To see this, we note that usually the transport current is determined by the fastest scattering rate which is the intravalley scattering. For the intravalley scattering with $f_0(\bm k)$ and $f_0(\bm k^\prime)$ from the same type of Weyl node, one can multiply  Eq.~\eqref{eq_bz} by $\mathcal{D}(\bm k)$ and integrate it around the Weyl node. The result is
\begin{align}\label{eq_cond}
&\frac{e}{\hbar}\int \frac{d\bm k}{8\pi^3}\bm E\cdot\left[ \bm vf_0^\prime- \bm \partial(\bm B\cdot \bm m f_0^\prime)+\frac{e}{\hbar}\bm B(\bm \Omega\cdot \bm v)f_0^\prime\right]=0\,.
\end{align}
However, for a Weyl node, although the integration of the first two terms vanish, the integration of the third term is proportional to the charge of the Weyl node and hence does not vanish. It indicates that charge can be pumped from the positive Weyl node to the negative Weyl node. This contradiction shows that the intervalley scattering is indispensable in solving the above Boltzmann equation.

To explicitly derive this solution, one can take $f_0(\bm k)$ to be from the positive Weyl node and $f_0(\bm k^\prime)$ from the negative Weyl node. Then under the relaxation time approximation, Eq.~\eqref{eq_cond} for the positive Weyl node becomes
\begin{align}\label{eq_ca2}
\frac{e^2}{\hbar^2}\int \frac{d\bm k}{8\pi^3}(\bm E\cdot\bm B)(\bm \Omega\cdot \bm v)f_0^\prime=\frac{\Delta N_+-\Delta N_-}{\tau_\text{int}}\,,
\end{align}
where $\Delta N_+$ and $\Delta N_-$ is the density of particles near the Fermi surface, and $\tau_\text{int}$ is the intervalley scattering time. Eq.~\eqref{eq_ca2} means that the pumping of particles from one Weyl node to another through drifting has to be balanced by the diffusive collision through the intervalley scattering process. It leads to an inhomogeneous chemical potential distribution in the momentum space
\begin{align}
\delta \mu_+-\delta\mu_-=-\frac{e^2}{4\pi^2\hbar^2}\frac{\tau_\text{int}}{g(\mu)}(\bm E\cdot \bm B)\,,
\end{align}
where $g(\mu)=\mu^2/(2\pi^2\hbar^3v^3)$ is the density of states at the Fermi surface and $\tau_\text{int}$ is the intervalley scattering time. This chemical potential difference for Weyl nodes with opposite chiralities is the chiral anomaly effect~\cite{Nielsen1983,Aji2012,Son2013,Burkov2015a,Armitage2018}.

This imbalance in the chemical potential can lead to a current through Eq.~\eqref{eq_cme}
\begin{equation}
\bm J_\text{CME}=\frac{e^4v^3\tau_\text{int}}{2\pi^3\hbar\mu^2}(\bm E\cdot \bm B)\bm B\,.
\end{equation}
This current is only present when $\bm E$ is parallel to $\bm B$.
The corresponding conductivity is quadratic in magnetic field and increases with it, indicating a negative longitudinal magnetoresistance. It corresponds to $\sigma_{(ij,kk),1}$ in Table.~1 and does not need to break time reversal or inversion symmetry.

Besides the semiclassical theory, the negative magnetoresistance has also been confirmed using chiral kinetic theory~\cite{Stephanov2012,Son2013a,Spivak2016,Hidaka2017,Sekine2017} and exact quantized Landau levels~\cite{Burkov2014}. In the semiclassical regime, the negative magnetoresistance for one pair of Weyl nodes is also extended in other situations. It has been shown that samples with Dirac points also have negative magnetoresistance, provided that the chiral charge is approximately conserved and the chiral relaxation time is large~\cite{Burkov2015}. Moreover, when the Dirac point opens a gap due to the suppression of the lattice symmetry, the chiral relaxation time is still present, so is the negative magnetoresistance~\cite{Andreev2018}.  Experimentally, the negative longitudinal magnetoresistance in Weyl and Dirac semimetals has been reported and interpreted as the chiral magnetic effect~\cite{Huang2015,Shekhar2015,Yang2015,Wang2016,Zhang2016,Kim2013,Xiong2015,Feng2015a,Li2015,Liang2014,Li2016,Zhang2017}, but similar signals could also arise from two different mechanisms discussed in the following two sections.

Finally, we comment that the Weyl nodes is gapless and near those band-crossing points, the semiclassical theory may fail. However, we can set the Fermi surface to be well above the Weyl point. The flux through the Fermi surface does not change with the Fermi energy. As a result, the discussion using the semiclassical theory still works.

\subsection{Intrinsic quadratic longitudinal magnetoresistance}
The magnetoconductivity due to chiral anomaly and chiral magnetic effect is one contribution to $\sigma_{(ij,kk)}$ in Eq.~\eqref{eq_jebb}. It is not a complete description of $\sigma_{(ij,kk)}$. To fully account for the drifting contribution to $\sigma_{(ij,kk)}$, one has to use the semiclassical theory up to second order, as presented in Sect.~III~(A). Specially, one can plug Eq.~\eqref{eq_f1} and \eqref{eq_dr} into Eq.~\eqref{eq_cur} and collect terms up to second order in magnetic field. The current responsible for $\sigma_{(ij,kk),1}$ is complicated but it can be put in several groups:
\begin{align}
\bm J=\bm J_1+\bm J_2+\bm J_3+\bm J_4+\bm J_5\,.
\end{align}

The first contribution has the form
\begin{align}
\bm J_1&=\frac{e^4}{\hbar^2}\tau\int \frac{d\bm k}{8\pi^3}[(\bm v\times \bm B)\cdot (\bm E\times \bm \Omega)](\bm v\times \bm B)\times \bm \Omega f_0^\prime\,.
\end{align}
This term is purely due to the anomalous velocity as both $\bm E\times \bm \Omega$ and $(\bm v\times \bm B)\times \bm \Omega$ are parts of the anomalous velocity. This current is particularly interesting. If the electromagnetic fields are along the same direction, which can be assume to be the $\hat{z}$ direction without loss of generality, this current becomes
\begin{align}
J_{1,z}&=-\frac{e^4}{\hbar^2}EB^2\tau\int \frac{d\bm k}{8\pi^3}(v_x\Omega_x+v_y\Omega_y)^2 f_0^\prime\,.
\end{align}
This will always lead to a negative longitudinal magnetoresistance.

The second contribution has the form
\begin{align}
\bm J_2=&\frac{e^2\tau}{2}\int \frac{d\bm k}{8\pi^3} \hat{e}_i\alpha_{ij}E_j(\bm B\cdot \bm m)^2 f_0^{\prime\prime}\notag\\
&-\frac{e^2\tau}{\hbar^2}\int \frac{d\bm k}{8\pi^3} \frac{\partial(\bm B\cdot \bm m)}{\partial \bm k} (\bm E\cdot \bm \partial_{\bm k})[(\bm B\cdot \bm m)f_0^\prime]\,.
\end{align}
This is solely due to the Zeeman coupling between magnetic field and the magnetic moment, containing both the spin and orbital part.

The third contribution has the form
\begin{align}
\bm J_3&=\frac{e^3}{\hbar^2}\tau \int \frac{d\bm k}{8\pi^3}(\bm E\cdot \bm B)\bm v (\bm \Omega\cdot \bm \partial_{\bm k})[(\bm B\cdot \bm m)f_0^\prime]\notag\\
&+\frac{e^3}{\hbar^2}\tau \int \frac{d\bm k}{8\pi^3}(\bm E\cdot \bm B)(\bm \Omega\cdot \bm v)\frac{\partial (\bm B\cdot \bm m)}{\partial \bm k}f_0^\prime\notag\\
&+\frac{e^3}{\hbar^2}\tau \int \frac{d\bm k}{8\pi^3}[(\bm v\times \bm B)\times \bm \Omega] (\bm E\cdot \bm \partial_{\bm k})[(\bm B\cdot \bm m)f_0^\prime]\notag\\
&+\frac{e^3}{\hbar^2}\tau \int \frac{d\bm k}{8\pi^3}\left\{\left[\frac{\partial (\bm B\cdot \bm m)}{\partial \bm k}\times \bm B\right]\times \bm \Omega\right\} (\bm E\cdot \bm v)f_0^\prime]\,.
\end{align}
This current is due to the coupling between the anomalous velocity and the Zeeman energy correction.

The fourth contribution has the form
\begin{align}
\bm J_4=&e^2\tau \int \frac{d\bm k}{8\pi^3}(2\hat{e}_i \alpha_{ij}E_jf_0^\prime+\bm v(\bm E\cdot \bm v) f_0^{\prime\prime})\varepsilon_{(2)}\notag\\
&+\frac{e^3\tau}{\hbar}\int \frac{d\bm k}{8\pi^3}[(\bm v\times \bm B)\cdot (\bm E\times \bm \Omega^\prime)]\bm v_0 f_0^\prime\notag\\
&-\frac{e^3\tau}{\hbar}\int \frac{d\bm k}{8\pi^3} (\bm v\cdot \bm \Omega^\prime)\bm B (\bm E\cdot \bm v) f_0^\prime\,.
\end{align}
The first line is due to the second order correction to the band energy and the remaining terms are due to the first order correction to the Berry curvature, i.e. $\bm \Omega^\prime$.

From the expressions of $\bm J_1$ to $\bm J_4$, we find that $\bm J_1$ is purely due to the orbital motion of Bloch electrons, while $\bm J_2$ to $\bm J_4$ can contain both spin and orbital contribution. The orbital part has been calculated in Ref.~\cite{Gao2017b} and the spin part has been calculated in Ref.~\cite{Dai2017}. Here $\bm J_1$ to $\bm J_4$ are complete expressions that contain the spin part, the orbital part, and the spin-orbital part contribution.

The last contribution is due to the correction to the chemical potential:
\begin{align}
\bm J_5&=\frac{e^2\tau}{\hbar} \int \frac{d\bm k}{8\pi^3} \bm v (\bm E\cdot \bm v)\left\{\mu_{(2)}f_0^{\prime\prime}-\frac{1}{2}[\mu_{(1)}]^2 f_0^{\prime\prime\prime}\right\}\,.
\end{align}

\begin{figure}[t]
  \centering
  \includegraphics[width=\columnwidth]{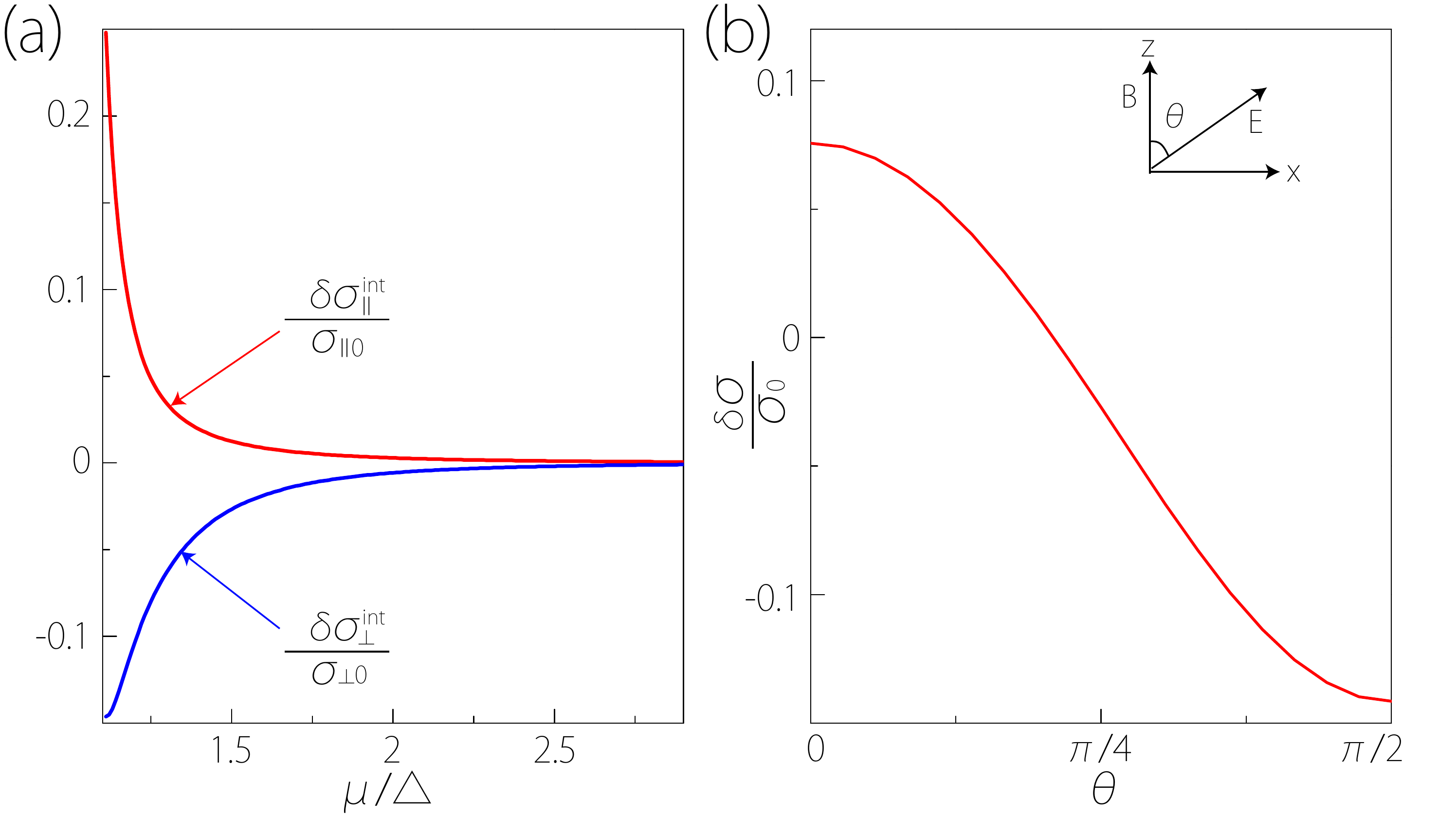}\\
  \caption{(a) Intrinsic magnetoconductivity versus the chemical potential. It is calculated for the model $\hat{H}=v(k_x\sigma_x+k_y\sigma_y)+(\Delta+k_z^2/2m)\sigma_z$, which can be realized by stacked Honeycomb lattices. Here $\sigma_{\perp 0}$ and $\sigma_{\parallel 0}$ are zero-magnetic-field conductivities under transverse and longitudinal configurations respectively. (b) Ratio between magnetoconductivity $\delta\sigma$
and zero-magnetic-field conductivity $\sigma_0$ versus the angle $\theta$ between $E$ and $B$ fields, as illustrated in the inset. Here the model parameters are chosen as
$B=2$T, $\Delta=50{\rm meV}$, $v_F=9.2\times 10^5 {\rm m/s}$, and $m^*=0.1m_e$ ($m_e$ is the free electron mass), and $\mu=60 {\rm meV}$ in (b). From Ref.~\cite{Gao2017b}.}\label{fig_nlmr}
\end{figure}

In nonmagnetic metals, $\sigma_{(ij,kk),3}$ is also nonzero. The corresponding current reads
\begin{align}
\bm J^\prime=-\frac{e^4}{\hbar^2}\tau^3\int \frac{d\bm k}{8\pi^3} \bm v[(\bm v\times \bm B\cdot \bm \partial_{\bm k})^2(\bm E\cdot \bm v)] f_0^\prime\,.
\end{align}
This current requires a peculiar Fermi surface such that the integration does not vanish~\cite{Pal2010}.

The current $\bm J$ and $\bm J^\prime$ complete the description of the quadratic magnetoresistance. In $\bm J^\prime$ the magnetic field always couples to the relaxation time. Therefore, the corresponding magnetoresistance follows Kohler's rule~\cite{Pippard1989}. In comparison, $\bm J$ violates Kohler's rule and its corresponding magnetoresistance does not depend on the relaxation time. Moreover, $\bm J$ does not need a highly anisotropic Fermi surface. As long as the Berry curvature, magnetic moment, and magnetic susceptibility does not vanish at the same time, $\bm J$ will be nonzero. This intrinsic contribution to magnetoresistance is also confirmed by a fully quantum mechanical treatment~\cite{Wang2018}.

The current $\bm J$ can yield a negative longitudinal magnetoresistance. In fact, for the Dirac Hamiltonian in Eq.~\eqref{eq_dirac}, the ratio of the quadratic longitudinal magnetoresistance $\delta \rho$ from $\bm J$ to that from chiral anomaly reads
\begin{align}
\frac{\delta \rho}{\delta \rho_\text{CA}}=+\frac{8}{45}\frac{\tau}{\tau_\text{int}}\,.
\end{align}
The global positive sign means that $\delta \rho$ is negative, like $\delta\rho_\text{CA}$. Usually $\tau\ll \tau_\text{int}$. Therefore, in Weyl semimetals, the negative magnetoresistance should be dominated by the contribution from the chiral anomaly in the semiclassical regime. However, the negative magnetoresistance from $\bm J$ can persist in metals without any Dirac/Weyl point, as shown in Fig.~\ref{fig_nlmr}. Panel (a) suggests that the longitudinal magnetoresistance is negative while the transverse one is positive. This causes the sign change of the effective magnetoresistance as the angle between $\bm E$ and $\bm B$ continuously changes from $0$ to $\pi/2$, as shown in Panel (b).

\begin{figure}
  \centering
  \includegraphics[width=\columnwidth]{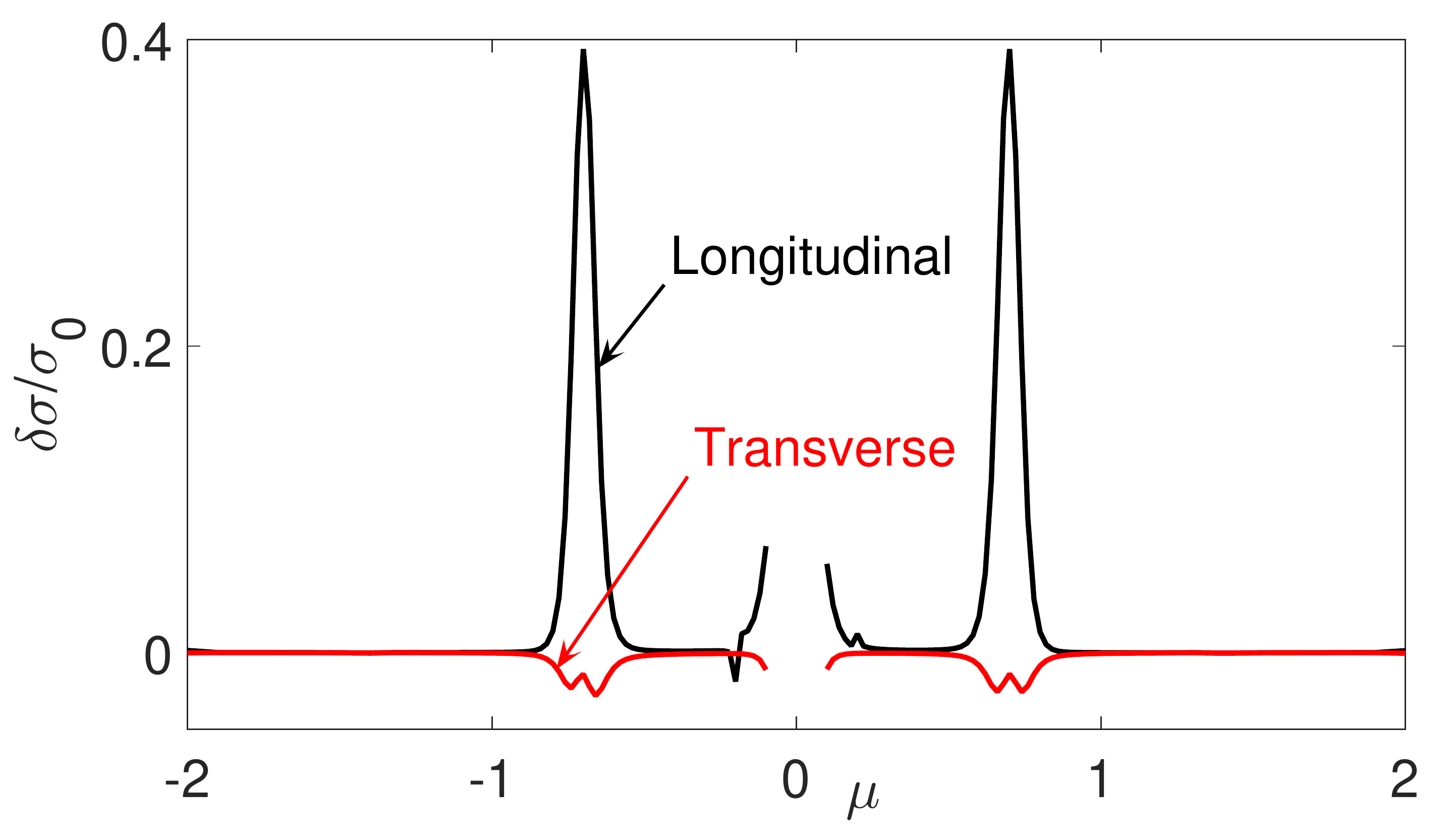}\\
  \caption{Transverse~(Red) and longitudinal~(Black) magnetoconductivity as a function of chemical potential. $\mu$ is in units of $\lambda$. For the red curve, $\sigma_{(xx,zz),1}\tau/\sigma_{xx}$ is plotted and for the black curve, $\sigma_{(zz,zz),1}\tau/\sigma_{zz}$ is plotted. The flux in one unit cell is $\phi=eBa^2/\hbar=3.79\times 10^{-3}$.}\label{fig_mr}
\end{figure}

It is interesting to further check the behaviour of the magnetoconductivity in a lattice model. We use the model Hamiltonian in Eq.~\eqref{eq_h0} and \eqref{eq_h1} and choose the parameters such that it is in the semiconductor phase as shown in Fig.~\ref{fig_eng0}~(d). We then calculate the quadratic magnetoconductivity for the transverse and longitudinal configuration, using the expression $\bm J_1$ to $\bm J_5$. Both spin and orbital part of the magnetic moment is considered. The result is presented in Fig.~\ref{fig_mr}. It can be found that both magnetoconductivity show strong dependence on the chemical potential. Consistent with the result in the low-energy model, the magnetoresistance will be strongly enhanced when chemical potential is near the band edge, as both the anomalous velocity and energy correction are enhanced. Moreover, the longitudinal magnetoconductivity is generally positive, indicating a negative longitudinal magnetoresistance, while the transverse magnetoconductivity is negative. Near the band edge, the longitudinal magnetoconductivity is larger than the transverse one, indicating a sizeable transition angle $\theta$ from negative to positive magnetoresistance, as the relative angle between $\bm E$ and $\bm B$ changes from $0$ to $\pi/2$. This is similar to the negative magnetoresistance induced by the chiral anomaly.

\subsection{Current Jetting}
\begin{figure}[t]
  \centering
  \includegraphics[width=\columnwidth]{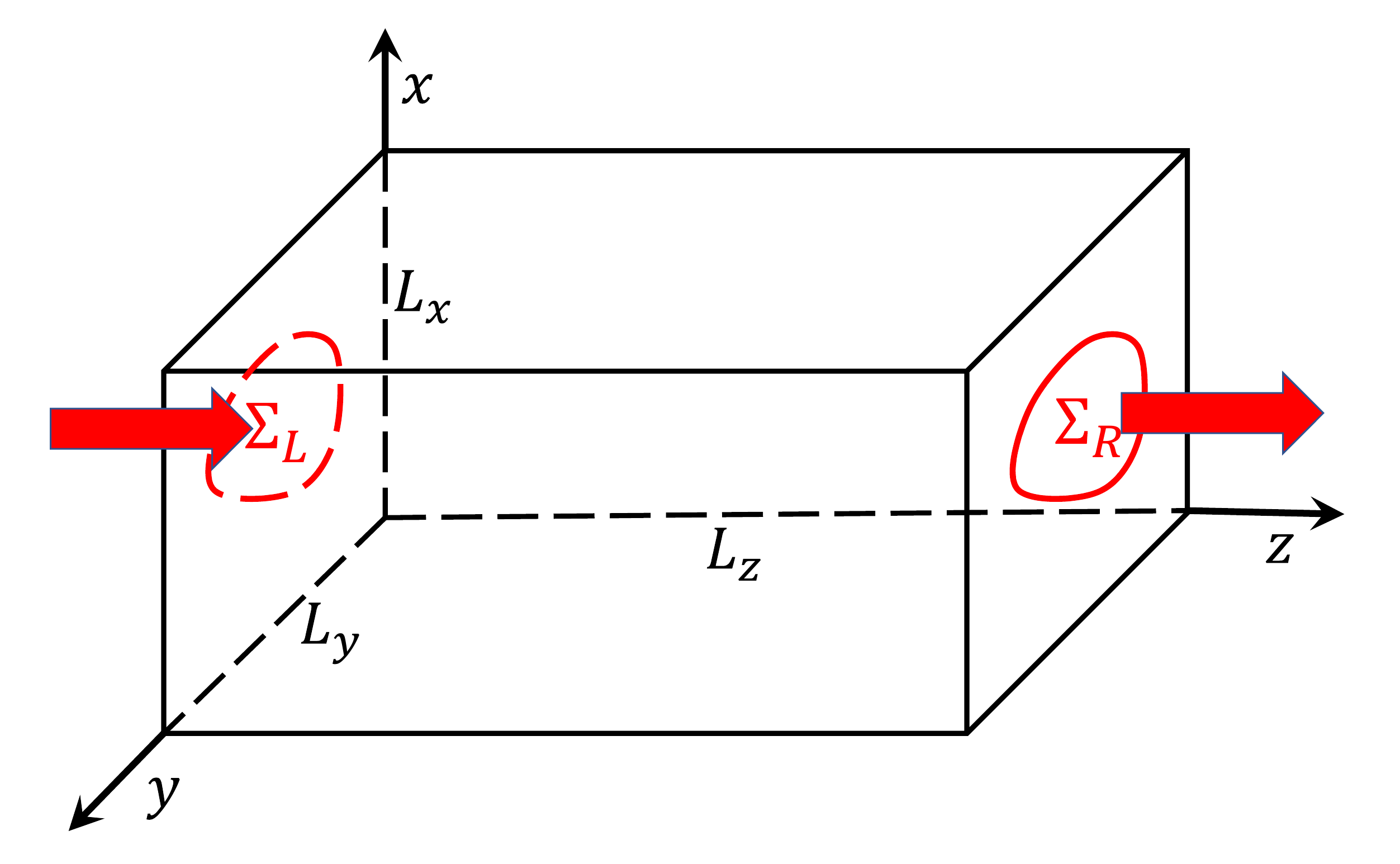}\\
  \caption{The current jetting problem. $\Sigma_L$ and $\Sigma_R$ stand for the regions of current input and output. $L_x$, $L_y$, and $L_z$ are the length of the sample along $x$, $y$, and $z$ directions.}\label{fig_cj}
\end{figure}

Beyond the Boltzmann transport theory, the magnetoresistance is also greatly affected when the current distribution in the sample is inhomogeneous. For example, for a matchstick sample with two leads for current input and output, as shown in Fig.~\ref{fig_cj}, the current in the sample can be highly inhomogeneous provided that the transverse magnetoresistance is sufficiently large~\cite{Yoshida1980,Pippard1989}. This is the current jetting problem.

The current jetting problem can be solved through the charge continuity equation: $\partial \rho/\partial t+\bm \nabla\cdot \bm J=0$. In the steady state $\partial \rho/\partial t=0$. Therefore, in the sample, the following equation for the electrostatic potential is satisfied
\begin{align}
\sigma_{ij}\partial_i\partial_j V=0\,.
\end{align}
Since the Hall conductivity is antisymmetric, it does not affect the above equation. However, it can affect the boundary condition. If the Hall conductivity can be ignored, and the longitudinal conductivity is determined by the Drude formula: $\sigma_{xx}=\sigma_{yy}=\sigma_0/(1+\omega_c^2\tau^2)$ and $\sigma_{zz}=\sigma_0$, the above equation can be put in the form of the Poisson equation
\begin{equation}\label{eq_pos}
\frac{\partial^2 V}{\partial x^2}+\frac{\partial^2 V}{\partial y^2}+R^2 \frac{\partial^2 V}{\partial z^2}=0\,,
\end{equation}
where $R=\sqrt{\sigma_{zz}/\sigma_{xx}}$. The boundary condition should also be determined by the charge continuity equation.

A large $R$ will greatly stretch the equipotential contour along $\hat{z}$ direction. It can cause a great portion of current lies in a sharp cone centered around the $\hat{z}$ axis~\cite{Pippard1989}. This is similar to the behaviour of the longitudinal magnetoresistance due to chiral anomaly. In fact, if the injection of the current only localized near a small region in the surface and does not spread to the whole surface, the current injection may cause a strong negative magnetoresistance~\cite{Huang2015, Reis2016,Yuan2016}.

\section{Outlook}
We have shown how to derive various nonlinear currents in the framework of the semiclassical theory and Boltzmann transport theory. Band properties beyond the spectrum play essential roles in those nonlinear currents. However, the semiclassical theory ignores the electron interaction and the modification to electron-impurity scattering from electromagnetic fields. The semiclassical theory has been extended in Fermi liquid using Keldysh formalism up to first order~\cite{Shindou2006,Shindou2008}. Moreover, the Boltzmann transport theory is also extended to treat the drifting and collision in equal-footing up to first order using the Wigner distriubtion function and quantum kinetic theory~\cite{Sekine2017,Culcer2017}. It is tempting to extend such theories up to second order to give a complete treatment of various nonlinear currents.

For the three nonlinear anomalous Hall effects, the one due to Berry curvature dipole has attracted several experimental efforts, while no experiments are reported to measure the other two corrections. The correction due to magnetic field should be generally present in any system and competes with the ordinary Hall effect. Distinguishing the magneto nonlinear Hall effect from the ordinary Hall effect is similar to extracting the intrinsic anomalous Hall from total anomalous Hall signal~\cite{Tian2009,Chun2007,Lee2004,Mathieu2004,Sales2006,Zeng2006}.

Moreover, in samples that have combined time reversal and inversion symmetry (but break each one separately), the linear anomalous Hall effect as well as the nonlinear Hall effect due to Berry curvature dipole vanishes identically and only the intrinsic nonlinear Hall current contributes. This type of materials is ideal to test the intrinsic nonlinear Hall effect. If the time reversal and inversion symmetry, as well as their combined symmetry are all broken, the anomalous Hall and its two electric nonlinear version should all be present. There are still methods to distinguish them. One can first perform a scaling of the relaxation time to distinguish the intrinsic contribution from the extrinsic contribution. This is similar to extracting intrinsic, skew-scattering and side-jump contribution from the total anomalous Hall signal~\cite{Tian2009,Chun2007,Lee2004,Mathieu2004,Sales2006,Zeng2006}. The nonlinear part can be further extracted by reversing the direction of the electric field and finding the difference in the Hall signal.

For the negative longitudinal magnetoresistance, the difficulty is to distinguish the contribution from chiral anomaly from the intrinsic and current jetting contributions. The current jetting effect may be reduced by attaching the lead across the whole end of the sample so that the current is uniformly injected. In comparison, the main difference between the chiral anomaly and the intrinsic contribution is a hierarchy of relaxation time. One needs to control the ratio of the intervalley or chiral relaxation time to the transport relaxation time to distinguish these two contributions.

There is another issue in evaluating nonlinear currents in first-principles codes. Depending on the type of the unperturbed Hamiltonian, the second order energy has different forms for Schrodinger Hamiltonian or effective tight-binding Hamiltonian, mainly due to different Hessian matrices. Then there are two methods to evaluate the second order energies and related effects: one can start from the Schr$\ddot{\rm o}$dinger Hamiltonian and use the wave function from first-principles codes as approximation to evaluate the analytical expression derived for the Schr$\ddot{\rm o}$dinger Hamiltonian; on the other hand, one can start from the effective tight-binding Hamiltonian and uses its response to electromagnetic fields to approach the real one. The second one requires to evaluate the effective Hessian matrices while that of the first one is simply $1/m\delta_{ij}$. It is tempting to study which method is more accurate.

\begin{acknowledgments}
We acknowledge useful discussions with Xiao Di.  This work is supported by the Department of Energy, Basic Energy Sciences, Grant No.~DE-SC0012509.
\end{acknowledgments}


%

\end{document}